\documentclass[journal]{IEEEtran}

\newcommand{\subFigureWidth}{0.4\textwidth}

\usepackage{fixltx2e}

\usepackage{cite}
\usepackage[noend]{algpseudocode}
\usepackage{algorithm}
\usepackage{booktabs}
\usepackage{graphicx}
\usepackage[caption=false]{subfig}
\usepackage[cmex10]{amsmath}
\usepackage{color}
\usepackage{array}
\usepackage{color}
\usepackage{multirow}
\usepackage{rotating}
\usepackage{mathdots}
\usepackage{framed}
\usepackage{amsmath}
\usepackage{comment}
\usepackage[usenames,dvipsnames,svgnames]{xcolor}

\usepackage{marginnote}
\usepackage{fancyhdr}

\alglanguage{pseudocode}
\algblockdefx[proc]{Proc}{EndProc}%
    [2]{{{\bf procedure} {\scshape #1}\hspace{0.2em}(\text{#2})}}%
    [0]{{{\bf end procedure}}}
\algblockdefx[init]{Init}{EndInit}
     [0]{{\bf initialization}}
     [0]{{\bf end initialization}}
\algblockdefx[upon]{Upon}{EndUpon}
     [1]{{\bf upon} #1}
     [0]{{\bf end upon}}
\renewcommand{\algorithmicindent}{2em}

\begin{document}

\clubpenalty=100000
\widowpenalty=10000

\newtheorem{definition}{Definition}
\newtheorem{theorem}{Theorem}
\newtheorem{lemma}{Lemma}

\title{ATLAS: \underline{A}daptive \underline{T}opology- and \underline{L}oad-\underline{A}ware \underline{S}cheduling}

\author{\IEEEauthorblockN{Jonathan Lutz, Charles J.\ Colbourn, and Violet R.\ Syrotiuk}
\IEEEauthorblockA{\\CIDSE, Arizona State University, Tempe, AZ ~~ 85287-8809}\\
Email: \{jlutz, colbourn, syrotiuk\}@asu.edu
}

\maketitle

\begin{abstract}
The largest strength of contention-based MAC protocols is simultaneously the largest weakness of their scheduled counterparts: the ability to adapt to changes in network conditions.
For scheduling to be competitive in mobile wireless networks, continuous adaptation must be addressed.
We propose ATLAS, an \underline{A}daptive \underline{T}opology- and \underline{L}oad-\underline{A}ware \underline{S}cheduling protocol to address this problem.
In ATLAS, each node employs a random schedule achieving its \emph{persistence}, the fraction of time a node is permitted to transmit, that is computed in a topology and load dependent manner.
A distributed auction (REACT) piggybacks offers and claims onto existing network traffic to compute a lexicographic max-min channel allocation.
A node's persistence $p$ is related to its allocation.
Its schedule achieving $p$ is updated where and when needed, without waiting for a frame boundary.
We study how ATLAS adapts to controlled changes in topology and load.
Our results show that ATLAS adapts to most network changes in less than 0.1s, with about 20\% relative error, scaling with network size. 
We further study ATLAS in more dynamic networks showing that it keeps up with changes in topology and load sufficient for TCP to sustain multi-hop flows, a struggle in IEEE 802.11 networks.
The stable performance of ATLAS supports the design of higher-layer services that inform, and are informed by, the underlying communication network.
\end{abstract}

\begin{IEEEkeywords}
Wireless networks, medium access control, adaptation.
\end{IEEEkeywords}


\fancypagestyle{firststyle}
{
   \renewcommand{\headrulewidth}{0pt}%
   \fancyhf{}
   \fancyfoot[C]{\footnotesize Submitted to IEEE Transactions on Mobile Computing -- \copyright 2013 IEEE}
}
\thispagestyle{firststyle}

\setlength{\heavyrulewidth}{0.12em}
\newcommand{\otoprule}{\midrule[\heavyrulewidth]}

\newcommand{\ie}{{\sl i.e.},{ }}
\newcommand{\eg}{{\sl e.g.},{ }}

\algrenewcommand{\algorithmiccomment}[1]{{\slshape{//{ }#1}}}

\newcommand{\CommentIndent}[1]           {\State\Comment{#1}\vspace{-0.0cm}}
\newcommand{\CommentNoIndent}[1]       {\State\hspace{-\algorithmicindent}\Comment{#1}\vspace{-0.0cm}}

\newcommand{\orLogic}	                           {\text{\bf { }or{ }}}
\newcommand{\andLogic}                         {\text{\bf { }and{ }}}
\newcommand{\true}                                {\text{\tt True}}
\newcommand{\false}                               {\text{\tt False}}
\newcommand{\send}[1]                           {\text{{\bf send} #1}}

\newcommand{\bidders}                           {\text{$D_j$}}
\newcommand{\auctions}                          {\text{$R_i$}}

\newcommand{\claim}                              {\text{\slshape claim}}
\newcommand{\offer}                               {\text{\slshape offer}}
\newcommand{\claimed}                           {\text{\slshape claimed}}

\newcommand{\maxClaim}                        {\text{$w_i$}}
\newcommand{\capacity}                          {\text{$c_j$}}
\newcommand{\weight}	                           {\text{\slshape weight}}



\newcommand{\biddersConst}                 {\text{$D_j^*$}}
\newcommand{\auctionsStable}               {\text{$A_{\text{stable}}$}}
\newcommand{\auctionsPlus}			{\text{$A^+$}}
\newcommand{\offerMin}                       {\text{$o_{\text{min}}$}}

\newcommand{\persistMin}                       {\text{$p_{\textup{min}}$}}
\newcommand{\persistDefault}                  {\text{$p_{\textup{default}}$}}
\newcommand{\nbrTimeOut}                    {\text{$t_{\textup{lostNbr}}$}}

\newcommand{\bidderClaims}[1]		{\text{{\slshape claims}[$#1$]}}
\newcommand{\bidderWeights}[1]     {\text{{\slshape weights}[$#1$]}}
\newcommand{\auctionOffers}[1]		{\text{{\slshape offers}[$#1$]}}

\newcommand{\auctionMsg}			{\text{\slshape auctionMsg}}
\newcommand{\bidderMsg}			{\text{\slshape bidderMsg}}

\newcommand{\bidderClaim}			{\text{\slshape claim}}
\newcommand{\auctionOffer}            {\text{\slshape offer}}
\newcommand{\available}			{\text{$\mathcal{A}_j$}}
\newcommand{\done}				{\text{\slshape done}}

\newcommand{\UpdateOffer}[1]  {\text{{\scshape UpdateOffer}\hspace{0.1em}(\text{#1})}}
\newcommand{\UpdateClaim}[1] {\text{{\scshape UpdateClaim}\hspace{0.1em}(\text{#1})}}


\section{Introduction}
\label{sec:intro}

Despite the well known shortcomings of IEEE 802.11 and other contention-based MAC protocols for mobile wireless networks---such as probabilistic delay guarantees, severe short-term unfairness, and poor performance at high load---they remain the access method of choice.
The primary reason is their ease in adapting to changes in network conditions, specifically to changes in topology and in load.
The lack of timely adaptation is the most serious limitation facing scheduled MAC protocols.
For scheduling to be competitive, continuous adaptation is required.

\emph{Topology-dependent} approaches to adaptation in scheduling alternate a contention phase with a scheduled phase.
In the contention phase, nodes exchange topology information used to compute a conflict-free schedule that is followed in the subsequent scheduled phase (see, as examples, \cite{ChlamtacP87,CZhu_1.01}). 
However, changes in topology and load do not always align with the phases of the algorithm resulting in a schedule that often lags behind the network state.

In contrast, the idea behind \emph{topology-transparent} scheduling is to design schedules independent of the detailed network topology \cite{IChlamtac_1.94,JJu_1.98}.
Specifically, the schedules do not depend on the identity of a node's neighbours, but rather on \emph{how many} of them are transmitting. 
Even if a node's neighbours change, its schedule does not; if the number of neighbours does not exceed the designed bound then the schedule guarantees success.
Though such schedules are robust to network conditions that deviate from the design parameters \cite{SyrotiukCY08}, because the schedules do not adapt, the technique remains a theoretical curiosity. 

In contention-based schemes, such as IEEE 802.11, a node computes \emph{implicitly} when to access the channel, basing its decisions on perceived channel contention.
We instead compute a node's \emph{persistence}---the fraction of time it is permitted to transmit---\emph{explicitly} in a way that tracks the current topology and load. 
To achieve this, we propose ATLAS, an \underline{A}daptive \underline{T}opology- and \underline{L}oad-\underline{A}ware \underline{S}cheduling protocol.
Channel allocation is a resource allocation problem where the demands correspond to transmitters, and the resources to receivers. 
ATLAS implements the \underline{R}\underline{E}source \underline{A}llo\underline{C}a\underline{T}ion computed by REACT, a distributed auction that runs {\em continuously}.
REACT piggybacks offers and claims onto existing network traffic to compute the lexicographic max-min allocation to transmitters which we call the \emph{TLA allocation}, emphasizing that it is both \emph{topology-} and \emph{load-aware}.
Each node's random schedule, achieving a persistence informed by its allocation, is updated 
whenever a change in topology or load results in a change in allocation.
While the slots of the schedule are grouped into frames, this is done only to reduce the variance in delay \cite{CColbourn_1.04}; there is no need to wait for a frame boundary to update the schedule.
Even though the random schedules may not be conflict-free, ATLAS is not contention-based; it does not select persistences or make scheduling decisions based on perceived channel contention---its decisions are based solely on topology and load.
We study how ATLAS adapts to controlled changes in topology and load, measuring convergence time, relative error, and scalability.
We also assess the ability of ATLAS to adapt in more dynamic network conditions.

To the best of our knowledge, ATLAS is the first scheduled MAC protocol able to adapt to changes in topology and load that is competitive with contention-based protocols in throughput and delay while realizing superior delay variance.
It achieves this through the continuous computation of the TLA allocation, and updating the schedule on-the-fly.
These updates occur only where and when needed.
By not requiring phases of execution and by computing persistences rather than conflict-free schedules, ATLAS eliminates the complexity of, and lag inherent in, topology-dependent approaches.
By not being dependent on the identity of neighbours, ATLAS shares the best of topology-transparent schemes (and also their potential for collisions) yet overcomes its weakness by being adaptive.
By not forcing updates to be frame synchronized, ATLAS shares the critical features of continuous adaptation with contention-based protocols.
As a result, ATLAS achieves predictable throughput and delay characteristics.
Such characteristics and information about localized capacity at the MAC layer may be used to inform higher layers, while end-to-end characteristics at higher layers may be used to inform ATLAS.
This may support the development of an agile, higher performing protocol stack.

The primary contributions of this paper are twofold:
(1) The REACT algorithm, an asynchronous, adaptive, and distributed auction that solves a general resource allocation problem to produce the TLA allocation.
(2) ATLAS, a MAC protocol that uses REACT to solve the specific problem of channel allocation in a wireless network where each node produces a random schedule with the number of transmission slots determined by its allocation.

The sequel is organized as follows:
Section~\ref{sec:adaptive_alg} defines a general resource allocation problem and presents the REACT algorithm, proving its correctness.
Section~\ref{sec:new_mac} expresses channel allocation as a resource allocation problem and defines ATLAS.
Related work is described in Section~\ref{sec:related_work}.
After describing the simulation set-up in Section~\ref{sec:set-up}, Section~\ref{sec:sim-results} studies how ATLAS adapts to controlled changes in topology and load, and to dynamic network conditions.
In Section~\ref{sec:discussion}, we discuss open issues and potential applications of REACT, including the design of higher-layer services that inform, and are informed by, the underlying communication channel.


\section{Distributed Resource Allocation --- REACT}
\label{sec:adaptive_alg}



We consider a general resource allocation problem.
Let $R$ be a set of $N$ {\em resources} with {\em capacity} $\boldsymbol{c} = (c_1,\ldots, c_N)$.
Let $D$ be a set of $M$ {\em demands} with magnitudes $\boldsymbol{w} = (w_1,\ldots,w_M)$.
Resource $j\in R$ is required by demands $D_j \subseteq D$.
Demand $i\in D$ consumes capacity at all resources in $R_i \subseteq R$ simultaneously. 
The {\em resource allocation} $\boldsymbol{s} = (s_1,\ldots,s_M)$, $s_i \geq 0$ defines the capacity reserved for the demands.
Resource allocation $\boldsymbol{s}$ is  {\em feasible} if $\sum_{i\in D_j} s_i \leq c_j$ for all $j \in R$ and $s_i \leq w_i$ for all 
$i\in D$.
Demand $i$ is {\em satisfied} if $s_i \geq w_i$.
Resource $j$ is {\em saturated} if  $\sum_{i\in D_j} s_i \geq c_j$.
Throughout, {\em capacity} refers to the magnitude of a resource.

\begin{definition}
\cite{MPioro_1.04} A feasible allocation $\boldsymbol{s}$ is {\em lexicographically max-min} if, for every demand $i\in D$,  either $i$ is satisfied, or there exists a saturated resource $j$ with $i \in D_j$ where  $s_i = \max ( s_k \colon k\in D_j )$. 
\label{def:maxmin}
\end{definition}


We now describe REACT, a distributed auction that computes the lexicographic max-min allocation.
In it, resources are represented by auctioneers and demands by bidders.
Each auctioneer maintains an \offer---the maximum capacity consumed by any adjacent bidder---and each bidder maintains a \claim---the capacity the bidder intends to consume at adjacent auctions.
The final claim of bidder $i$ defines allocation $s_i$.
Auctioneer $j$ satisfies Def.~\ref{def:maxmin} locally by increasing its offer in an attempt to become saturated while maintaining a feasible allocation.
Bidder $i$ satisfies Def.~\ref{def:maxmin} locally for demand $i$ by increasing its claim until it is satisfied or has a maximal claim at an adjacent auction.
Through continuous updates of offers and claims, the auctioneers and bidders eventually converge on the lexicographic max-min allocation. 
We give precise definitions of auction and bidder behaviour next.

Bidder $i$ knows \maxClaim{} and maintains set \auctions.
Offers are stored in \auctionOffers{}; \auctionOffers{j} holds the offer last received from auctioneer $j$.
Bidder $i$ constrains its \claim{} to be no larger than \maxClaim{} or the smallest offer from auctioneers in \auctions,
\begin{equation}
\bidderClaim = 
          \min \left( \{\auctionOffers{j}: j\in \auctions\}, \maxClaim \right).
\label{eq:bidderClaim}
\end{equation}

Auctioneer $j$ knows \capacity{} and maintains set \bidders.
Bidder claims are stored in \bidderClaims{}; \bidderClaims{i} holds the claim last received from bidder $i$.
Auctioneer $j$ identifies set $\biddersConst \subseteq \bidders$ containing bidders with claims strictly smaller than its offer,
\begin{equation}
\biddersConst = \{b : b\in \bidders, \bidderClaims{b} < \auctionOffer\}.
\label{eq:auctionBcnstrnd}
\end{equation}
Bidders in \biddersConst{} are either satisfied or are constrained by another auction and cannot increase their claims in response to a larger offer from auctioneer $j$.
Bidders in $\bidders \setminus \biddersConst$ are constrained by auction $j$.
They may increase their claims in response to a larger offer.
Resources left unclaimed by bidders in \biddersConst{},
\begin{equation}
\available = \capacity - \left(\textstyle\sum_{i\in \biddersConst} \bidderClaims{i}\right),
\label{eq:auctionAvailable}
\end{equation}
remain available to be offered in equal portions to bidders in $\bidders \setminus \biddersConst$.
If claims of all bidders in \bidders{} are smaller than the offer (\ie $\bidders = \biddersConst$), there are no bidders to share the available resources in \available.
The auctioneer sets its offer to \available{} plus the largest claim, ensuring that any bidder in \bidders{} can increase its claim to consume resources in \available:
\begin{equation} 
\auctionOffer 
= \begin{cases}
\available / |\bidders \setminus \biddersConst|, 
& \!\!\text{if $\bidders \ne\biddersConst$,} \\
\available + \max \left( \bidderClaims{i}: i\in \bidders\right),
& \!\!\text{otherwise.}
\end{cases}
\label{eq:auctionOffer}
\end{equation}
\begin{algorithm}[tbp]
\begin{algorithmic}[1]
\footnotesize

\Upon {initialization}
  \State $\auctions \gets \emptyset$
  \State $\maxClaim \gets 0$
  \State \UpdateClaim{}
\EndUpon

\vspace{0.1cm}
\Upon {receiving a new demand magnitude $w_i$}
  \State \UpdateClaim{}
\EndUpon

\vspace{0.1cm}
\Upon {receiving  \offer{} from auctioneer $j$}
  \State $\auctionOffers{j} \gets \offer$
  \Comment{Remember the offer of auctioneer $j$.}
  \State \UpdateClaim{}
\EndUpon

\vspace{0.1cm}
\Upon {bidder $i$ joining auction $j$}
  \State $\auctions \gets \auctions \cup j$
  \Comment{Resource $j$ is now required by demand $i$.}
  \State \UpdateClaim{}
\EndUpon

\vspace{0.1cm}
\Upon {bidder $i$ leaving auction $j$}
  \State $\auctions \gets \auctions \setminus j$
  \Comment{Resource $j$ is no longer required by demand $i$.}
  \State \UpdateClaim{}
\EndUpon

\vspace{0.1cm}
\Proc{UpdateClaim}{}

  \State\Comment{Select the claim to be no larger than the smallest offer or \maxClaim.}
  \State $\bidderClaim \gets 
          \min \left( \{\auctionOffers{j}: j\in \auctions\}, \maxClaim \right)$
  \label{alg:line:bidderClaim}

  \State \send{\claim} to all auctions in $R_i$
  \label{alg:line:send}

\EndProc
\end{algorithmic}
\caption{REACT Bidder for Demand $i$.}
\label{alg:bidder}
\end{algorithm}

\begin{algorithm}[tbp]
\begin{algorithmic}[1]
\footnotesize

\Upon {initialization}
  \State $\bidders \gets \emptyset$
  \State $\capacity \gets 0$
  \State \UpdateOffer{}
\EndUpon

\vspace{0.1cm}
\Upon {receiving a new capacity of \capacity}
  \State \UpdateOffer{}
\EndUpon

\vspace{0.1cm}
\Upon {receiving \claim{} from bidder $i$}
  \label{alg:line:recv_claim}
  \State $\bidderClaims{i} \gets \claim$
  \Comment{Remember the claim of bidder $i$.}

  \State \UpdateOffer{}
\EndUpon

\vspace{0.1cm}
\Upon {bidder $i$ joining auction $j$}
  \State $\bidders \gets \bidders \cup i$
  \Comment{Demand $i$ now requires resource $j$.}
  \State \UpdateOffer{}
\EndUpon

\vspace{0.1cm}
\Upon {bidder $i$ leaving auction $j$}
  \State $\bidders \gets \bidders \setminus i$
  \Comment{Demand $i$ no longer requires resource $j$.}
  \State \UpdateOffer{}
\EndUpon

\vspace{0.1cm}
\Proc{UpdateOffer}{}

  \State $\biddersConst \gets \emptyset$
  \State $\available \gets \capacity$
  \State $\done \gets \false$ 
  \While{$(\;\done = \false \;)$}
    \State\Comment{If \biddersConst{} contains all bidders in \bidders, then auction $j$ does not}
    \State\Comment{constrain any of the bidders in \bidders.}
    \If {$(\; \biddersConst = \bidders \;)$}
      \State $\done \gets \true$
      \State $\offer \gets \available + \max \left( \{\bidderClaims{i}: i\in \bidders\}\right)$

    \CommentNoIndent{Otherwise, auction $j$ constrains at least one bidder in \bidders.}
    \Else
      \State $\done \gets \true$
      \CommentIndent{What remains available is offered in equal portions to the}
      \CommentIndent{bidders constrained by auction $j$.}
      \State $\auctionOffer \gets \available / |\bidders \setminus \biddersConst|$
      \CommentIndent{Construct \biddersConst{} and compute \available{} for the new offer.}
      \ForAll{$b \in \{\bidders \setminus \biddersConst \}$}
          \If {$(\;\bidderClaims{b} < \auctionOffer \;)$}
            \State $\biddersConst \gets \biddersConst \cup b$
            \State $\available \gets \available - \bidderClaims{b}$
            \State $\done \gets \false$
          \EndIf
      \EndFor
    \EndIf
   \EndWhile

    \State \send{\offer} to all bidders in $D_j$

\EndProc
\end{algorithmic}
\caption{REACT Auctioneer for Resource $j$.}
\label{alg:auction}
\end{algorithm}

Alg. \ref{alg:bidder} and Alg. \ref{alg:auction} describe actions taken by the bidders and auctioneers of REACT in response to externally triggered events.
Collectively, auctioneers and bidders know the inputs to the allocation problem and bidder claims converge on the lexicographic max-min allocation; the claim of bidder $i$ converges on $s_i$.

The correctness of Alg.~\ref{alg:bidder} and Alg.~\ref{alg:auction} is established in two steps: 
Lemma~\ref{lemma:alg_correctness} establishes forward progress on the number of auctioneers to have converged on their final offer.
Theorem~\ref{thm:alg_correctness} employs Lemma~\ref{lemma:alg_correctness} to show eventual convergence to the lexicographic max-min allocation.
Let $\claim_i$ denote the claim of bidder $i$ and $\offer_j$ the offer of auctioneer $j$.
Assume that the resource allocation remains constant for the period of analysis, that bidder $i$ knows \auctions{} and \maxClaim, and that auctioneer $j$ knows \bidders{} and \capacity.
Further assume communication between adjacent auctioneers and bidders is not delayed indefinitely.
A claim or offer is {\em stable} if it has converged on its final value.
Denote by \auctionsStable{} the set of auctioneers whose offers are stable and remain the smallest among all offers.

\begin{lemma}\label{lemma:alg_correctness}
Suppose \auctionsStable{} contains $k$ auctioneers, $0\leq k <N$.
Then, within finite time, at least one auctioneer converges on the next smallest offer \offerMin{}.
Offers equal to \offerMin{} are stable and remain smaller than all other offers not in \auctionsStable.
\end{lemma}

\begin{IEEEproof}
Wait sufficient time for every bidder $i$ to send a new claim to auctioneers in $R_i$ and for every auctioneer $j$ to send a new offer to bidders in $D_j$.
Let \offerMin{} be the smallest offer of an auctioneer not in \auctionsStable{}.
Assume to the contrary that $\offer_x$ for some $x\notin \auctionsStable$ is the {\em first} to become smaller than \offerMin.
By Eq.~\ref{eq:auctionBcnstrnd} and \ref{eq:auctionOffer}, a decrease to $\offer_x$ can only occur {\em after} a bidder $y$ at auction $x$ with $\claim_y < \offer_x$ increases its claim.
By Eq.~\ref{eq:bidderClaim}, $\claim_y$ can increase only {\em after} its limiting constraint starts out smaller than $\offer_x$ and increases.
Constraints in the system smaller than $\offer_x$ are maximum claims, offers from \auctionsStable{}, and offers equal to \offerMin.
Maximum claims and offers from \auctionsStable{} do not change, leaving some $x'$ with $\offer_{x'}=\offerMin$ as the only potential limiting constraint for $\claim_y$.
By Eq.~\ref{eq:auctionBcnstrnd} and \ref{eq:auctionOffer}, $\offer_{x'}$ can increase only {\em after} one of its bidders $y'$ reduces its claim to be smaller than $\offer_{x'}$.
By Eq.~\ref{eq:bidderClaim}, $\claim_{y'}$ can get smaller only {\em after} one of its auctioneers, say $x''$, reduces its offer to be $\offer_{x''}<\offerMin=\offer_{x'}$ contradicting the assumption that $\offer_x$ is the {\em first} to become smaller than \offerMin.
Therefore, offers equal to \offerMin{} remain smaller than offers not from \auctionsStable.

By Eq.~\ref{eq:auctionBcnstrnd} and \ref{eq:auctionOffer}, any $j$ offering \offerMin{} can change only after a bidder $i$ at auction $j$ with $\claim_i \leq \offerMin$ changes.
By Eq.~\ref{eq:bidderClaim}, $\claim_i$ only changes if its limiting constraint changes.
Potential limiting constraints include \maxClaim, offers from \auctionsStable{}, and offers equal to \offerMin.
These constraints are stable; therefore, offers equal to \offerMin{} are stable.
\end{IEEEproof}

\begin{theorem}\label{thm:alg_correctness}
Bidders and auctioneers of Alg.~\ref{alg:bidder} and Alg.~\ref{alg:auction} compute the lexicographic max-min allocation.
\end{theorem}

\begin{IEEEproof}
We apply Lemma~\ref{lemma:alg_correctness} to show by induction that every auctioneer eventually computes a stable offer.

{\em Base Case}: 
Consider an allocation problem with arbitrary $\maxClaim$, $\capacity$, $R_i$, and $D_j$ for $1 \leq i \leq M$, $1\leq j \leq N$.
Let $|\auctionsStable| = 0$.
By Lemma~\ref{lemma:alg_correctness}, at least one auctioneer eventually converges on a smallest offer \offerMin. 
Offers equal to \offerMin{} are stable and remain smallest among all offers.
Add auctioneers offering \offerMin{} to \auctionsStable{}; $|\auctionsStable|\geq 1$.

{\em Inductive Step}: 
Let $|\auctionsStable| = k$, $1\leq k <N$. 
Then, by Lemma~\ref{lemma:alg_correctness} a non-empty set of auctioneers \auctionsPlus{} with $\auctionsPlus{} \cap \auctionsStable = \emptyset$ eventually converge on the next smallest offer.
Offers from \auctionsPlus{} remain smaller than offers not from \auctionsPlus{} or \auctionsStable{} and are stable.
Add \auctionsPlus{} to \auctionsStable{}; $|\auctionsStable| \geq k+1$.

By induction, all auctioneers are eventually added to \auctionsStable{}.
Wait for auctioneers to send their offers to adjacent bidders.
Bidder claims are now stable.
By Eq.~\ref{eq:bidderClaim}, bidder $i$ is either satisfied with its claim ($\claim_i = \maxClaim$) or its claim is maximal at an auction in $R_i$.
By Definition~\ref{def:maxmin}, the claims are lexicographic max-min.
\end{IEEEproof}




\section{The ATLAS MAC Protocol}
\label{sec:new_mac}

Channel allocation in wireless networks can be expressed as a resource allocation problem.
In this context,  transmitters correspond to the demands in $D$ and receivers to the resources in $R$.
Label transmitters $\{1,\dots,M\}$ and receivers $\{1,\dots,N\}$.
A transmitter with a non-zero demand magnitude is {\em active}.
Receiver $j$ is in $R_i$ if it is within transmission range of transmitter $i$ and transmitter $i$ is active.
$D_j$ contains the active transmitters for which receiver $j$ is within transmission range.
Receiver $j$ is {\em adjacent} to transmitter $i$ if $j\in R_i$ and $i\in D_j$.
The sets \bidders{} and \auctions{} capture the network topology for active transmitters.
For load, \maxClaim{} is set to the percentage of slots required to support the demand at transmitter $i$.
Transmitters with no demand (\ie $\maxClaim=0$) receive an allocation of zero slots: they are not active.
Receiver capacities are set to one, targeting 100\% channel allocation.
The lexicographic max-min solution $\boldsymbol{s}=(s_1,\ldots, s_M)$ for a given topology and traffic load is the {\em TLA allocation}.

To apply REACT to channel allocation, we integrate it into ATLAS, a simple random scheduled MAC protocol.
Although REACT could instead augment contention-based schemes, we choose to work within a scheduled environment, a traditionally difficult setting for adaptation.
In ATLAS, each node runs a REACT bidder (Alg. \ref{alg:bidder}) and a REACT auctioneer (Alg. \ref{alg:auction}) continuously.
Auctioneers and bidders discover each other as they hear from one another and rely on the host node to detect lost adjacencies.
The network topology is {\em implicit} in the sets \auctions{} and \bidders{}.
Each node updates its bidder's demand magnitude to accurately reflect its traffic load.
Offers and claims are encoded using eight bits each and are embedded within the MAC header of all transmissions to be piggybacked on existing network traffic.
The encoding supports a total of 256 values for offers, claims, and persistences uniformly distributed between 0 and 1; the error in the representation does not exceed 0.004.
Adding fields for an offer and claim to data packets and acknowledgements results in a communication overhead of four bytes per packet.
For the slot size and data rate simulated in Section~\ref{sec:sim-results}, the overhead is 0.36\%.
A node's offer and claim are eventually received by all single-hop neighbours reaching the bidders and auctioneers that need to know the offer and claim.
In time, the bidder claims in REACT converge on the TLA allocation $\boldsymbol{s}$.


\begin{figure}
\centering
\resizebox{0.9\linewidth}{!}{
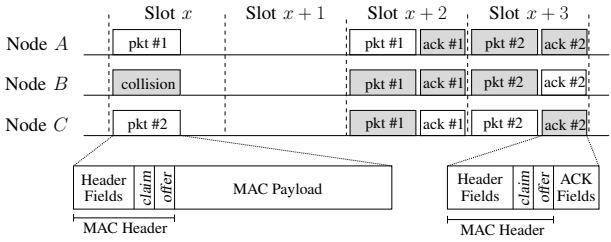
}
\caption{
Example transmissions in ATLAS of two packets in a network of three fully connected nodes.
The first packet is sent from node $A$ to node $C$. 
The second packet is sent from node $C$ to node $B$.
Transmissions are coloured white and receptions are shaded grey.
The frame structure is shown for a data packet and an acknowledgement.
}
\label{fig:ex_slots}
\end{figure}

Packets are acknowledged within the slot they are transmitted and slots are sized accordingly.
Unacknowledged MAC packets are retransmitted up to ten times before they are dropped by the sender.
\figurename~\ref{fig:ex_slots} shows that collisions are possible in ATLAS, and that successful transmissions are acknowledged in the same slot.
The transmissions collide in slot $x$; they are repeated (successfully) in slots $x+2$ and $x+3$.
\figurename~\ref{fig:ex_slots} also shows the frame structure.

The TLA allocation can be interpreted directly as a set of persistences in a $p$-persistent MAC \cite{ATanenbaum_1.03}.
However, we achieve lower variation in delay by introducing the notion of a frame \cite{CColbourn_1.04}. 
Specifically, ATLAS divides time into slots which are organized into frames of $v$ slots.
Node $i$ operates at persistence $p_i=s_i$.
At the start of every frame and upon any change to $p_i$, node $i$ computes $k_i = \lfloor p_i v \rfloor+1$ with probability $\pi_i$ and $k_i = \lfloor p_i v \rfloor$ with probability $1-\pi_i$ where $\pi_i=p_iv-\lfloor p_iv \rfloor$.
Node $i$ constructs a transmission schedule of $k_i$ slots selected uniformly at random.
Over many frames, $E[k_i]/v$ equals $p_i$ where $E[k_i]$ is the expectation for $k_i$.

\begin{figure}
\centering
\resizebox{0.9\linewidth}{!}{
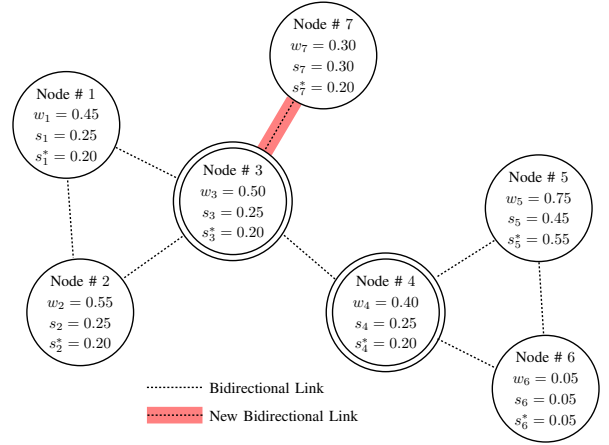
}
\caption{
Example network showing the TLA allocation computed by REACT before and after an added link in the topology.
$w_i$ identifies a node's demand, $s_i$ its initial TLA allocation, and $s_i^*$ its TLA allocation after the added link.
Resource capacities are set to one.
Double-lined circles identify nodes with saturated resources.
}
\label{fig:ex_alloc}
\end{figure}

\figurename~\ref{fig:ex_alloc} shows the TLA allocation in a small example network before and after a change in topology.
Node 7 starts out disconnected from the other nodes and moves within range of node 3.
In REACT, node 3 starts out offering 0.25 which is claimed by the bidders of nodes 1, 2, 3, and 4.
With the claims of node 3 and 4 limited by the offer of node 3 and the claim of node 6 limited by its demand, the auctioneer at node 4 is free to offer 0.45, which is claimed by node 5.
Upon detecting node 7 as a neighbour, the auctioneer at node 3 decreases its offer to 0.20.
The bidders at nodes 1, 2, 3, 4, and 7 respond by reducing their claims accordingly.
The smaller claims of the bidders at nodes 3 and 4 allow the auctioneer at node 4 to increase its offer to 0.55. 
The bidder at node 5 responds by increasing its claim to 0.55.
It can be verified that, before and after the topology change, the claims of the bidders (\ie the values of $s_i$ and $s_i^*$) are lexicographically max-min; that is, every claim is satisfied or is maximal at an adjacent auction.
Consider the topology with node 3 and node 7 connected.
The bidder at node 6 is satisfied.
The bidders at nodes 1, 2, 3, 4, and 7 are maximal at the auction of node 3.
The bidder at node 5 is maximal at the auction of node 4.



There are many implementation choices to be made in applying REACT to channel allocation.
We identify three binary choices---lazy or eager persistences, physical layer or MAC layer receivers, and weighted or non-weighted bidders---and three configurable parameters---\persistMin, \persistDefault, and \nbrTimeOut.
The choices are described here; they are evaluated in Section~\ref{sec:sim-results}.


\subsection{Lazy or Eager Persistences}
\label{sec:lazy_eager}


A lazy approach sets persistence $p_i$ equal to the \claim{} of bidder $i$.
Once converged, 
$p_i$ matches the TLA allocation interpreted as a persistence.
There is a potential disadvantage with being lazy.
For many applications, nodes cannot predict future demand for the channel; they can only estimate demand based on past events, \ie packet arrival rate or queue depth.
As a consequence, \maxClaim{} lags the true magnitude of the demand at node $i$. 
If \maxClaim{} is the limiting constraint for the \claim{} of bidder $i$, $p_i$ can be sluggish in response to increases in demand.
Alternatively, an eager approach sets persistence $p_i = \min \left( \auctionOffers{j}: j\in \auctions \right)$, breaking the direct dependence on \maxClaim.
Under stable conditions, a node's channel {\em occupancy}, the fraction of time it spends transmitting, matches its TLA allocation; its occupancy is limited by the availability of packets to transmit which is no larger than \maxClaim, even when $p_i > \maxClaim$.
By allowing $p_i>\maxClaim$, the persistence is made more responsive to sudden increases in demand.


\subsection{Physical Layer or MAC Layer Receivers}
\label{sec:phy_mac}

A central objective of the TLA allocation is to ensure that no receiver is overrun. 
In a wireless network, receivers can be defined in terms of physical layer or MAC layer communication.
At the physical layer, every node is a receiver.
At the MAC layer, packets are filtered by destination address; a node is only a receiver if one of its neighbours has MAC packets destined to it.
MAC layer receivers can increase channel allocation by over-allocating at non-receiving nodes.
However, the overallocation can slow detection of new receivers.
Physical receivers prevent overallocation at {\em any} receiver, making the allocation more responsive to changes in traffic where nodes become receivers.

%

\subsection{Weighted or Non-Weighted Bidders}
\label{sec:weighted}


We have described a MAC protocol where transmitters are represented by equally weighted bidders.
For applications requiring multiple demands per transmitter, \ie nodes servicing more than one traffic flow, we propose the {\em weighted} TLA allocation.
The demands of weighted bidders are comprised of one or more {\em demand fragments}; the number of fragments accumulated into a demand is the demand's {\em weight}.
Let $\gamma_i$ be the weight for demand $i$.
Demand fragments in demand $i$ have magnitude $w_i/\gamma_i$.
The weighted TLA allocation defines the lexicographically max-min vector $u=(u_1,\ldots, u_N)$ where 
$u_i$ is the allocation to {\em each} demand fragment in demand $i$ for a total allocation of $u_i\gamma_i$ to demand $i$.
REACT can be extended to compute the weighted TLA allocation.
To do this, each bidder must inform adjacent auctions of its weight.
Sixteen unique weights (with a four-bit representation) may be sufficient for many applications.


\subsection{Minimum Persistence \persistMin}
\label{sec:min_persist}

A node can maintain a persistence of zero without impacting the communication requirements of its bidder.
For auctioneers, a persistence of zero is problematic.
If a receiver becomes overwhelmed by neighbouring transmitters, a non-zero persistence is needed to quiet the neighbours.
To accomplish this, the node enforces a minimum persistence \persistMin, creating dummy packets if necessary, whenever the sum of claims from adjacent bidders exceeds the auction capacity.

\subsection{Overriding the TLA Allocation with \persistDefault}
\label{sec:default_persist}

There are two conditions where a node constrains its persistence to be no larger than \persistDefault.
The first is when it has no neighbours.
While the TLA allocation permits an isolated node to consume 100\% of the channel, it cannot discover new neighbours if it does so.
The second time a node employs \persistDefault{} is for a short period after the discovery of a new neighbour.
It is possible for several nodes operating with large persistences to join a neighbourhood at about the same time.
If the persistences are large enough, neighbour discovery can be hindered.
For both scenarios, limiting the persistence to \persistDefault{} facilitates efficient neighbour discovery.


\subsection{Adaptation to Topology Changes and \nbrTimeOut{}}
\label{sec:nbrtimeout}

Changes in network topology are detected externally to REACT.
In ATLAS, neighbour discovery is performed independently by each node. 
If a node hears from a new neighbour, then the node notifies its bidder of the new auction and its auctioneer of the new bidder.
Conversely, if a node has not heard from a neighbour in more than \nbrTimeOut{} seconds, it presumes the node is no longer a neighbour and informs its auctioneer and bidder accordingly.

\section{Related Work}
\label{sec:related_work}

This paper focuses on the TLA allocation, its continuous distributed computation, and its application to setting transmitter persistences.
In this section, we review a representative set of scheduled MAC protocols, observing how each selects a node's persistence and adapts to topology and load.

Any finite schedule used in a cyclically repeated way can be generalized as a $(k,v)$-schedule with $k$ transmission slots per frame of $v$ slots, producing an effective persistence of $p=k/v$.
Examples include the random schedules of \cite{CColbourn_1.04,JLutz_01.10} where each node selects its $k$ transmission slots randomly from the set of $v$ slots in the frame.
Topology transparent schemes \cite{IChlamtac_1.94,JJu_1.98,SyrotiukCY08} also implement $(k,v)$-schedules.
These schedules rely on only two design parameters: $N$, the number of nodes in the network, and $D_\text{max}$, the maximum supported neighbourhood size.
These schedules guarantee each node a collision-free transmission opportunity from each of its neighbours at least once per frame, provided the node's neighbourhood size does not exceed $D_\text{max}$.
$(k,v)$-schedules do not adapt to variations in neighbourhood size or traffic load.
The combinatorial requirements for variable-weight topology transparent schedules (variable $k$) are explored in \cite{JLutz_02.12}, but no construction nor protocol using them is given.

A class of topology-dependent scheduled protocols compute distance-2 vertex colourings of the network graph to achieve TDMA schedules with spatial reuse.
The colourings assign one transmission slot to each node and do not adapt to traffic load.
One of the first distributed protocols to bound the number of colours is proposed in \cite{ChlamtacP87}.
Distributed-RAND (DRAND) \cite{RheeWMX09} is a distributed implementation of RAND (a centralized algorithm for distance-2 colouring \cite{Ramanathan97}).
DRAND runs a series of loosely synchronized rounds.
A colour is assigned in each round to one or more nodes in different two-hop neighbourhoods.
DRAND is employed by Zebra-MAC (Z-MAC) \cite{RheeWAMS12} to compute schedules over which to run CSMA/CA.
Nodes are given priority access to their own slot, but also allowed to contend for access in other unused slots, as is done in \cite{IChlamtac_01.99}.
Due to the complexity of DRAND, schedules are only computed once during network initialization.

Other topology-dependent schemes support variable persistences.
The periodic slot chains proposed in \cite{JGentian_1.12} are not limited to the structure of a fixed length frame and can support variable and arbitrarily precise persistences.
A slot chain is defined by its starting transmission slot and period between its consecutive transmission slots.
By combining multiple slot chains with different periods, schedules are constructed targeting any rational persistence in the range $[0,1]$.
The computation of slot chains provided in \cite{JGentian_1.12} is centralized; a distributed mechanism to adaptively compute the slot chains remains an open problem.
In \cite{CZhu_1.01}, a five phase reservation protocol (FPRP) computes conflict-free schedules where a node can reserve one or more transmission slots in the frame to achieve variable persistences. 
Reservation frames are run periodically rather than on a demand basis and, therefore, may not accommodate the current topology and traffic load.

In SEEDEX \cite{RRozovsky_1.01}, nodes do not attempt to derive conflict-free schedules. 
They learn the identities of their two-hop neighbours and adjust transmission probabilities (\ie persistences) to improve the likelihood of collision-free transmissions.
The transmission probabilities accommodate the number and identity of neighbours, but not traffic load.

In our earlier work \cite{JLutz_01.12}, a distributed algorithm for computing the TLA allocation is provided; however, the algorithm assumes a fixed topology and does not adapt to changes in the network.
REACT solves these limitations by asynchronously adapting to changes in both topology and traffic demand.


\section{Simulation Set-up}
\label{sec:set-up}

We now describe the simulations used to produce the experimental results presented in Section~\ref{sec:sim-results}.
Table \ref{tbl:atlas_configs} lists the four ATLAS configurations simulated.
The {\em Nominal} configuration employs eager persistences, defines receivers in terms of MAC layer communication, and operates with unweighted bidders.
The other three configurations differ from the Nominal case by a single choice and are named accordingly.

\begin{table}
\renewcommand{\tabcolsep}{0.15cm}
\centering
\caption{ATLAS configurations selected for simulation.}
\begin{tabular}{cccc}
\hline
{\bf Configuration } &
{\bf Eager (0) } &
{\bf MAC (0) or  }  & 
{\bf Unweighted (0) }  \\ 
{\bf Name} &
{\bf or Lazy (1) } &
{\bf Physical (1) } &
{\bf or Weighted (1) }  \\ 
\hline
Nominal   & 0  & 0 & 0 \\
Lazy Persistences & 1 & 0 & 0 \\
Physical Receivers  & 0 & 1 & 0 \\
Weighted Bidders  & 0 & 0 & 1 \\
\hline
\end{tabular}
\label{tbl:atlas_configs}
\end{table}

\subsection{Scenario Details}
\label{sec:scenarios}

Unless otherwise noted, all four configurations run with $\persistDefault=0.05$, $\nbrTimeOut=0.5$s, and $\persistMin=0.01$.
The selection of \persistDefault{} and \nbrTimeOut{} are justified by results in Figs.~\ref{fig:boot_conv_defaultp}, \ref{fig:error_top_rwp_timeout}, and \ref{fig:thput_top_rwp_timeout}.
The selection of \persistMin{} is based on \cite{JLutz_01.12}.
Frames contain $v=100$ slots of length 800$\mu$s (1100 bytes per slot).
Simulations are run using the {\tt ns-2} simulator \cite{ns2}.
Each wireless node is equipped with a single half-duplex transceiver and omni-directional antenna whose physical properties match those of the 914 MHz Lucent WaveLAN DSS radio.
The  data rate for all simulations is 11 Mbps.
The transmission and carrier sense ranges are 250m.

Each simulation runs a network scenario composed of a randomly generated topology and a randomly generated traffic load.
Unless specified otherwise, topologies contain 50 randomly placed nodes constrained to a $300\times1500\text{m}^2$ area.
With the exception of the multi-hop TCP flows in Section~\ref{sec:results_tcp}, each traffic load consists of single-hop constant rate traffic.
Four traffic loads are simulated: 20\% and 80\% of nodes loaded with small demands ($75\pm50$ pkts/s), 20\% and 80\% of nodes loaded with large demands ($500\pm50$ pkts/s). 
Nodes loaded with traffic are selected at random and the demand magnitudes are selected uniformly at random from the specified range.
The packet destination is selected dynamically from the set of neighbouring nodes as the packet is passed down to the MAC layer.
For the Weighted Bidders configuration, each demand is assigned a random integer weight between one and five.
Traffic is generated by constant bit rate generators and transported over UDP; packets are 900 bytes in length, leaving room in each slot for header bytes and a MAC layer acknowledgement.
Combined with the random placement of nodes and the addition of mobility, these four traffic loads enable simulation of a wide variety of network conditions.

\subsection{Relative Error}
\label{sec:rel-error}

A metric of interest is the average relative error for a node's persistence with respect to the TLA allocation.
Error is reported in two parts: relative excess and deficit persistence error.
Errors are measured per node over 80ms consecutive intervals in time (equal to the length of one MAC frame).
We compute the average relative excess error and average relative deficit error for a given sample set of persistence measurements.
The relative errors are ratios, requiring use of the geometric rather than arithmetic mean.
But, the errors are often zero, preventing direct use of their mean.
Instead, we convert errors into accuracies eliminating zeros from the data set for a more meaningful geometric average.
The average relative accuracies are converted back to relative errors.


\section{Evaluation of ATLAS} \label{sec:sim-results}

Results from \cite{JLutz_01.12} show the TLA allocation applied in a static network to maintain expected delay and throughput compared to IEEE 802.11, while reducing the variance for both metrics. 
The TLA allocation nearly eliminates packets dropped by the MAC layer.
In this section, we build on these results, focusing on the efficient distributed computation of the TLA allocation in the face of changes in topology and load.
The results presented here work to answer four questions:
\begin{enumerate}
\item Can ATLAS converge quickly on the TLA allocation? 
\item Can ATLAS scale to larger networks?
\item Can ATLAS keep up with changes in a mobile network?
\item Can ATLAS adapt to multi-hop traffic flows? 
\end{enumerate}
The first question is addressed Sections~\ref{sec:results_conv_net_init},~\ref{sec:results_conv_dmnd_chg},~and~\ref{sec:results_conv_top_chg}.
The second is addressed in Section~\ref{sec:results_scalability}.
The third and fourth questions are addressed in Sections~\ref{sec:results_mobility} and \ref{sec:results_tcp}, respectively.
Continuing the focus on adaptation, Section~\ref{sec:mac_comparison} provides comparisons with several scheduled protocols.



\subsection{Convergence after Network Initialization}
\label{sec:results_conv_net_init}

\begin{figure}[t]
\centering
\includegraphics[width=\subFigureWidth]
{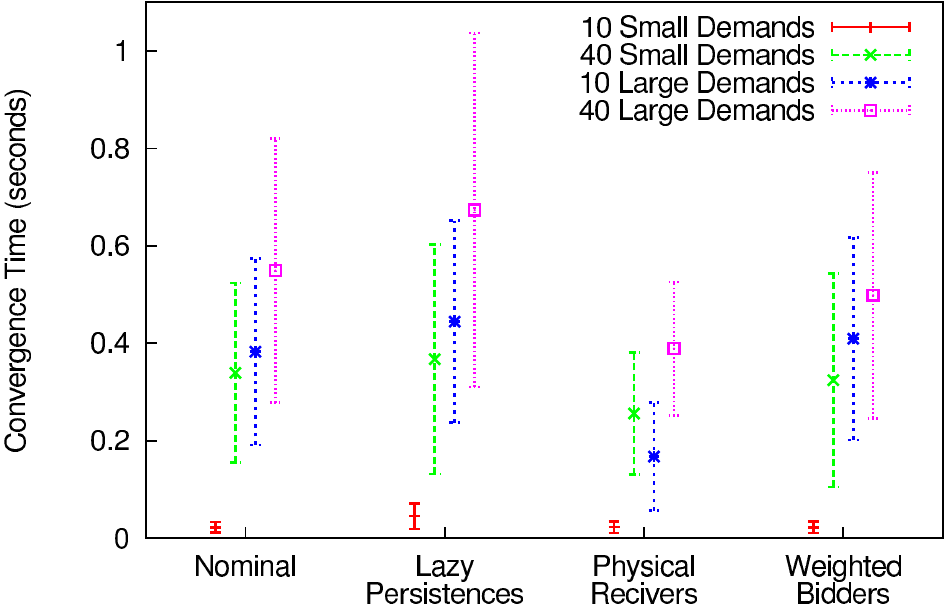}
\caption{
Convergence time following network initialization.
}
\label{fig:boot_conv}
\end{figure}

\figurename~\ref{fig:boot_conv} reports average convergence times for all four ATLAS configurations.
Error bars denote the arithmetic standard deviation from the mean for each sample set.
Convergence is measured from network initialization (time = 0) to the time ATLAS converges on the TLA allocation.
Times are collected from simulations of 1000 network scenarios simulated four times each, once per configuration.
There are 250 scenarios for each traffic load.

The Physical Receivers configuration converges fastest in less than 0.4s on average for networks with 40 large demands and faster for other traffic loads.
The extra step of detecting MAC receivers slows convergence.
The Lazy Persistences configuration is the slowest with an average convergence time of 0.67s for networks with 40 large demands. 
The strict limit on persistences enforced by this configuration slows convergence compared to the others.

\begin{figure*}%
\centering
	\subfloat[Relative excess and deficit persistence errors.]{
		\includegraphics[width=\subFigureWidth]%
		{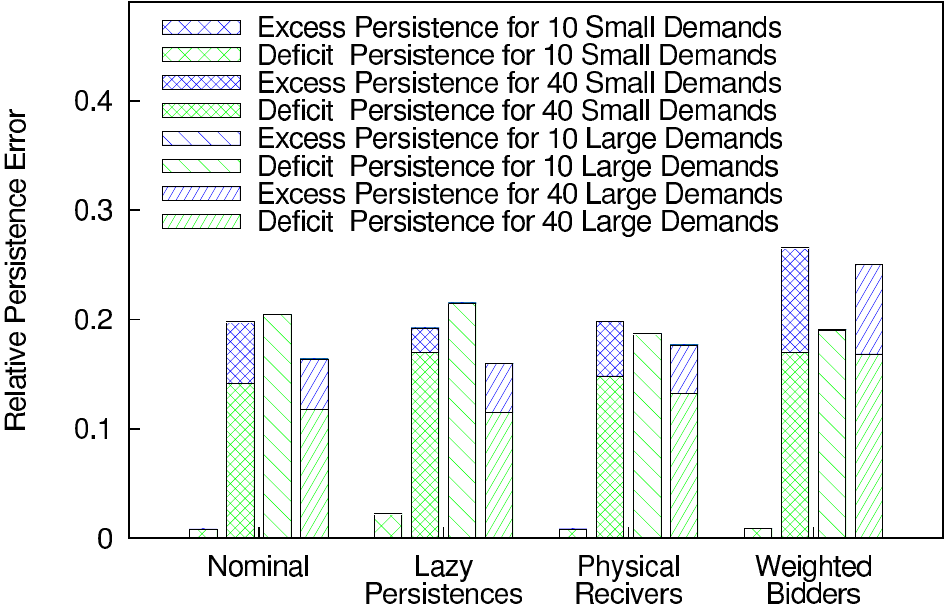}%
		\label{fig:boot_error}
	}
	\hspace{0.65in}
	\subfloat[Convergence time vs. error for the Nominal configuration.]{
		\includegraphics[width=\subFigureWidth]%
		{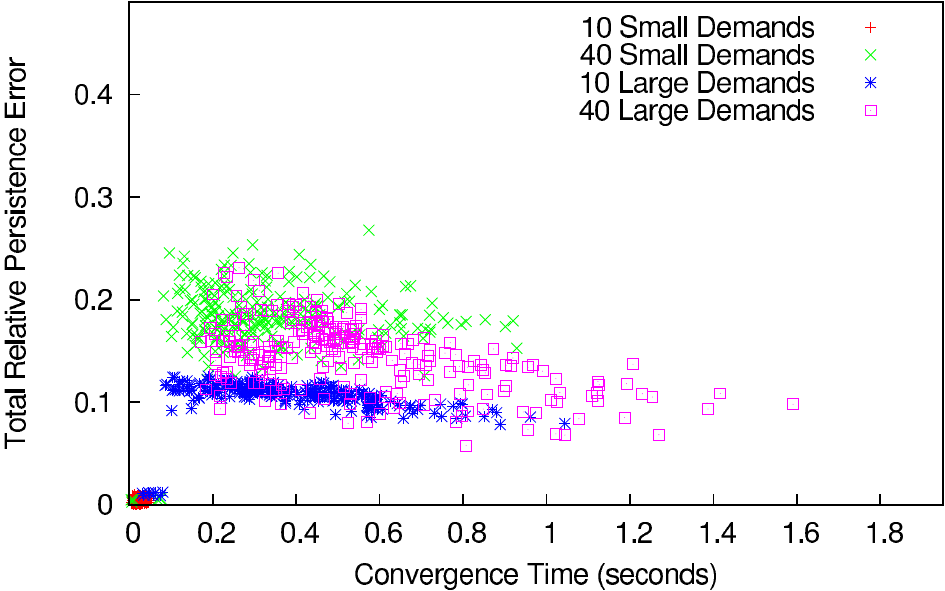}%
		\label{fig:boot_error_vs_conv}
	}
	\caption{Relative persistence error for ATLAS.}
	\label{fig:boot_error_both}
\end{figure*}

\figurename~\ref{fig:boot_error} shows average excess and deficit relative persistence errors for all four configurations.
The averages are computed for nodes with a non-zero TLA allocation and only during convergence.
Nodes are observed to operate within approximately 20\% of their TLA allocation regardless of configuration.
Deficit errors are larger than excess errors reflecting a tendency to converge from below, rather than above, the TLA allocation.
In \figurename~\ref{fig:boot_error_vs_conv},
each data point reflects the convergence time ($x$-coordinate) and total relative persistence error ($y$-coordinate) for one simulation of the nominal configuration.
The data shows relative persistence error to be fairly consistent from network to network with a maximum observed error of 27\%.

\begin{figure}[t]
\centering
\includegraphics[width=\subFigureWidth]
{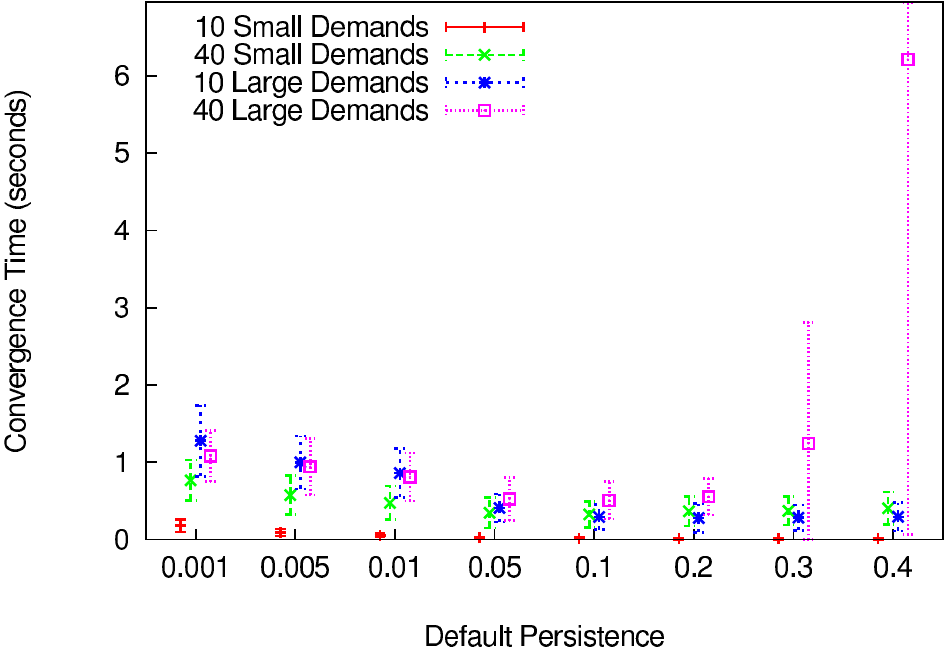}
\caption{
Convergence times when run with varying default persistences. 
}
\label{fig:boot_conv_defaultp}
\end{figure}

\figurename~\ref{fig:boot_conv_defaultp} reports convergence time for the Nominal configuration while varying \persistDefault.
Convergence is measured for simulations of 1000 network scenarios, 250 of each traffic load.
The scenarios are simulated eight times each, once per default persistence: 0.001, 0.005, 0.01, 0.05, 0.1, 0.2, 0.3, and 0.4. 
Small default persistences ($\persistDefault\leq 0.01$) limit a node's ability to communicate during neighbour discovery, slowing convergence.
Large default persistences ($\persistDefault\geq 0.3$) permit nodes to transmit with large persistences before they discover their neighbours.
In networks with 40 large demands, the large persistences can overwhelm the channel preventing neighbour discovery and delaying convergence.
ATLAS is robust to the selection of \persistDefault{} with a suitable range of $[0.05$--$0.2]$.
For the remaining simulations, \persistDefault{} is 0.05.

\subsection{Convergence after a Change in Demand}
\label{sec:results_conv_dmnd_chg}

\begin{figure*}[ht!]%
\centering
	\subfloat[Convergence time after a demand change.]{
		\includegraphics[width=\subFigureWidth]%
		{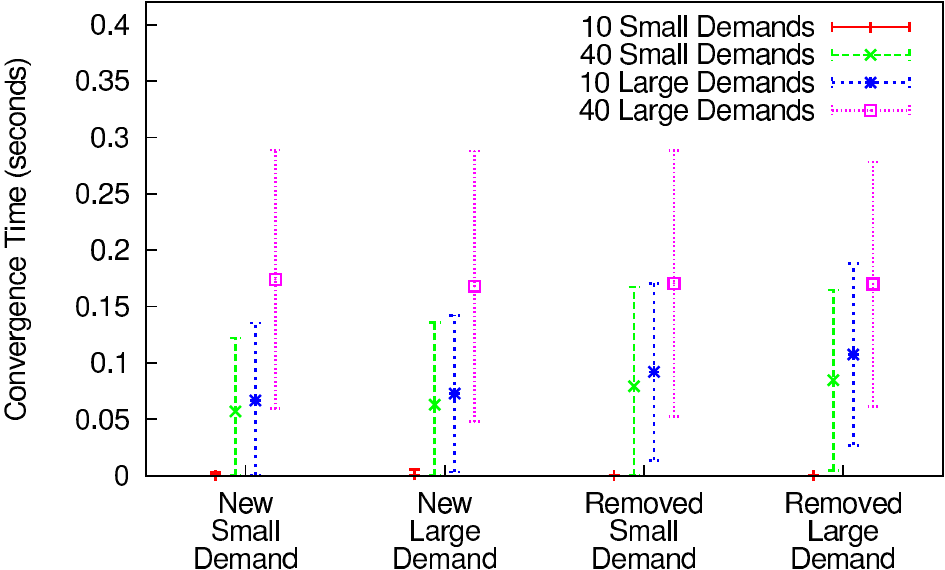}%
		\label{fig:dmnd_chg_conv}
	}
	\hspace{0.65in}
	\subfloat[Relative persistence error after a demand change.]{
		\includegraphics[width=\subFigureWidth]%
		{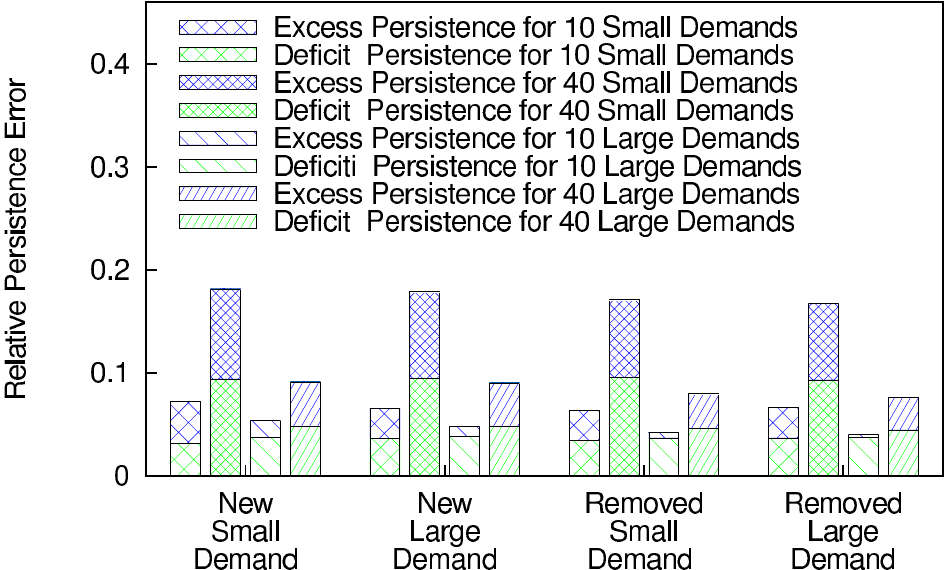}%
		\label{fig:dmnd_chg_error}
	}
	\caption{Convergence time and relative persistence error during convergence following a single demand change.}
	\label{fig:dmnd_chg}
\end{figure*}

\figurename~\ref{fig:dmnd_chg} reports convergence times and relative persistence errors for the Nominal configuration following a change to a single demand magnitude.
Four types of demand are simulated: a new small demand, a new large demand, a removed small demand, and a removed large demand.
New demands start with magnitude zero and change to $75\pm50$ pkts/s for small demands and to $500\pm50$ pkts/s for large demands.
Removed small demands and removed large demands start at $75\pm50$ pkts/s and at $500\pm50$ pkts/s, respectively; both change to zero.
The four demand change types are simulated under the four traffic loads. 
REACT is allowed to converge on the initial TLA allocation prior to the demand change.
Convergence times and error measurements are taken from simulations of 4000 network scenarios, 250 for each of the 16 demand change and traffic load combinations.

\figurename~\ref{fig:dmnd_chg_conv} reports convergence times measured from the time of the change to the time of convergence on the new TLA allocation.
The largest convergence times of approximately 0.175s are found in networks loaded with 40 demands.
The average convergence time for the other scenarios is 0.125s or smaller.
\figurename~\ref{fig:dmnd_chg_error} shows relative persistence errors measured during convergence at nodes whose TLA allocation are affected by the demand change.
Persistences are observed to be within 10\% of the TLA allocation.

\subsection{Convergence after a Change in Topology}
\label{sec:results_conv_top_chg}

\begin{figure*}%
\centering
	\subfloat[Convergence time after a topology change.]{
		\includegraphics[width=\subFigureWidth]%
		{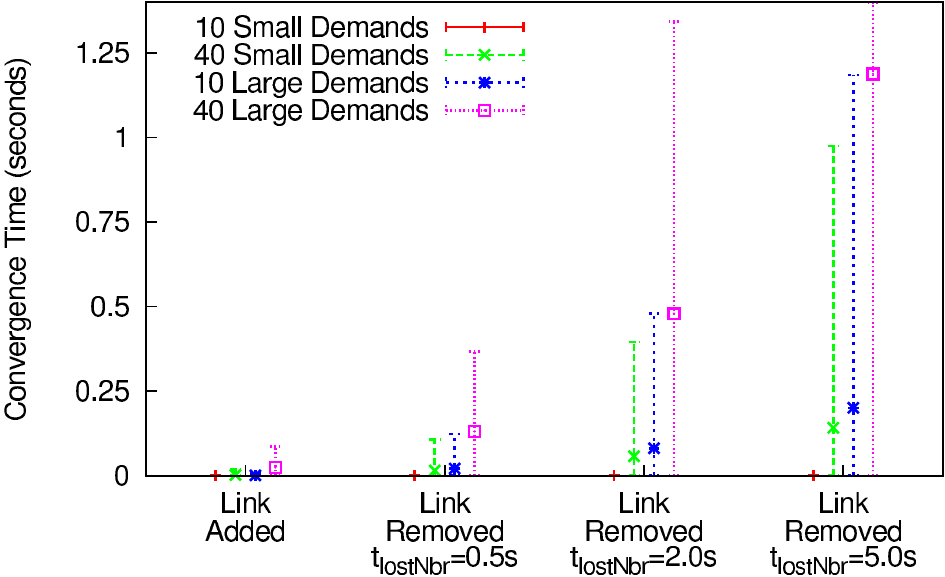}%
		\label{fig:top_chg_conv}
	}
	\hspace{0.65in}
	\subfloat[Relative persistence error after a topology change.]{
		\includegraphics[width=\subFigureWidth]%
		{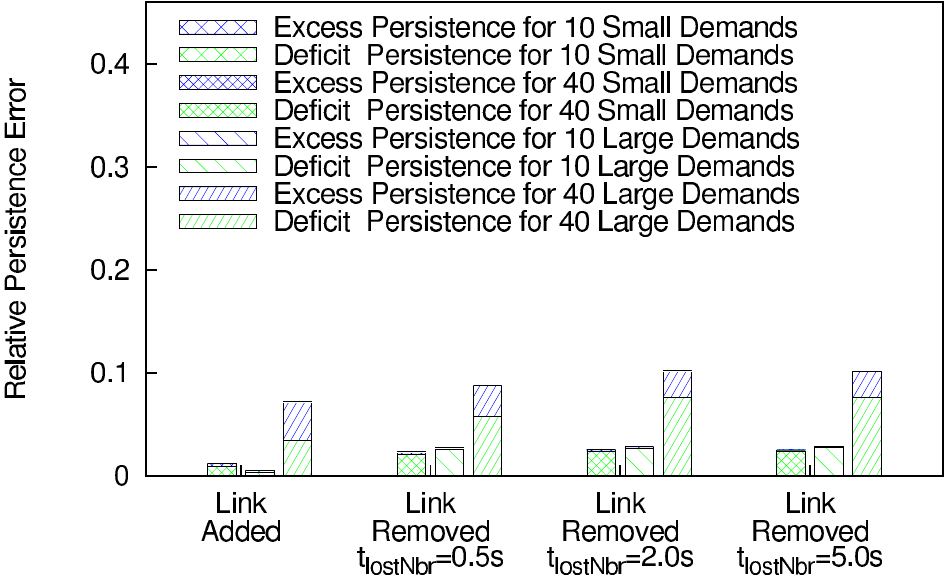}%
		\label{fig:top_chg_error}
	}
	\caption{Convergence time and relative persistence error following a single topology change.}
	\label{fig:top_chg_conv_error}
\end{figure*}

\figurename~\ref{fig:top_chg_conv_error} reports convergence time and relative persistence error following two types of topology change: the creation of a link and the removal of a link between a pair of nodes.
Simulations are run on 2000 network scenarios, 250 for each topology change type and traffic load combination.
Networks that lose a link are simulated once per neighbour timeout \nbrTimeOut{} of 0.5s, 2.0s and 5.0s.

Network topologies are generated as follows.
A first node is placed at a random location in the simulation area.
For topologies gaining a link, a second node is placed just {\em outside} the transmission range of the first node with a trajectory {\em toward} the first node.
For topologies losing a link, the second node is placed just {\em inside} the transmission range of the first node with a trajectory {\em away} from the first node.
The remaining 48 nodes are placed at random locations in the simulation area.
The distance travelled by the second node is constrained to avoid unintentional topology changes.

The expected convergence time following the addition of a new link is 0.025s.
For \nbrTimeOut$=$0.5s, convergence is reached in less than 0.13s on average.
For \nbrTimeOut$=$2.0s and \nbrTimeOut$=$5.0s, the large convergence times are dominated by \nbrTimeOut{}.
Except for the simulations of \figurename~\ref{fig:timeout_error_thput}, all others configure ATLAS with \nbrTimeOut$=$0.5s.
During convergence, nodes affected by the topology change are observed to operate within 4\% of their TLA allocation on average.
These numbers are striking.
The small convergence times stem from a counterintuitive feature of the TLA allocation: the majority of topology changes do not affect the TLA allocation.
A new link only has an effect if the link connects a bidder with an auction that lacks the capacity to support the bidder's claim.
Even in heavily loaded networks, many auctions have spare capacity to support a new bidder.
For these scenarios, convergence is instantaneous.

\subsection{Scalability to Large Networks}
\label{sec:results_scalability}

\begin{figure}
\centering
\includegraphics[width=\subFigureWidth]
{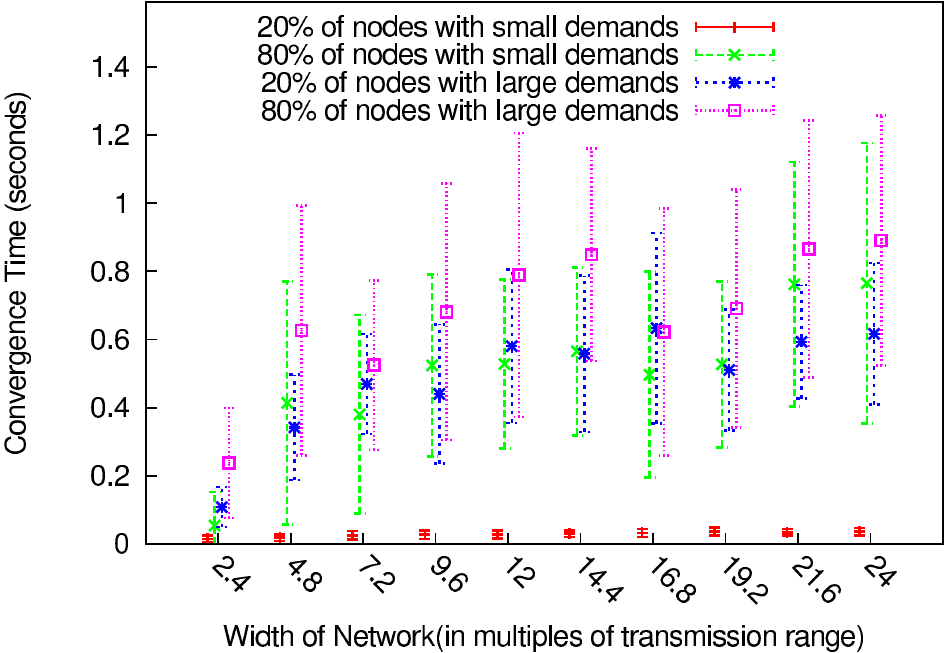}
\caption{
Convergence times as the width of the network grows.
}
\label{fig:conv_multihop}
\end{figure}

We now turn to results demonstrating ATLAS's scalability.
We simulate 10 network sizes with the $x$-dimension ranging from 600m (2.4 hops) to 6000m (24 hops) in 600m increments; the $y$-dimension is held constant at 300m.
The number of nodes is selected to keep the average neighbourhood density constant across all network sizes.
\figurename~\ref{fig:conv_multihop} reports convergence times for 4000 network scenarios, 100 of each traffic load and network size combination.
The convergence of ATLAS in large networks is striking.
In networks spanning 24 hops, convergence is reached in an average of 0.89s, a mere 40\% increase compared to networks spanning 4.8 hops.



\begin{figure}
\centering
\includegraphics[width=\subFigureWidth]
{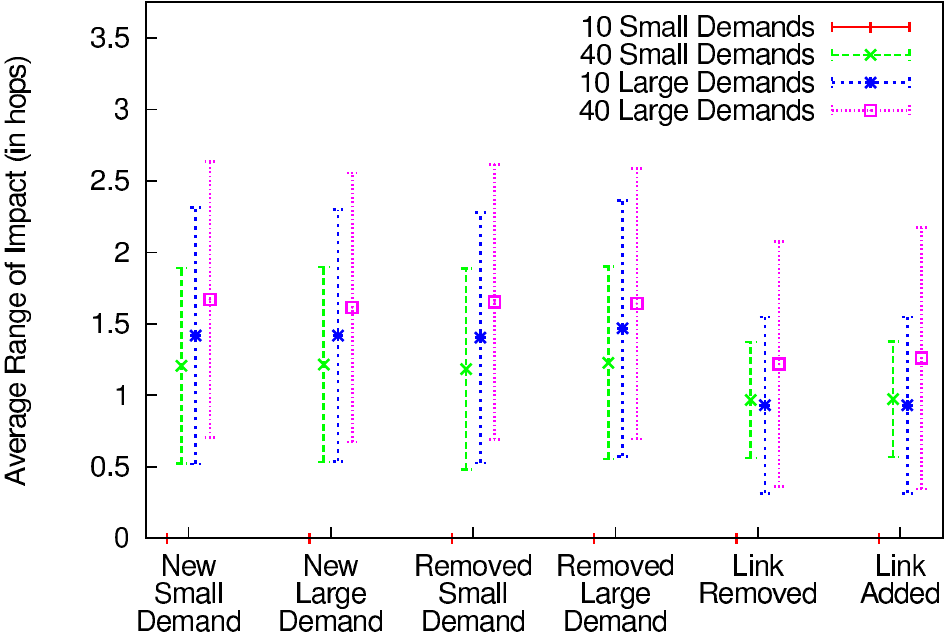}
\caption{
Average range of impact (in hops) for a demand or topology change.
}
\label{fig:range_impact}
\end{figure}

The impressive convergence times, particularly those of networks spanning 12 or more hops, suggest that convergence happens locally, allowing distant neighbourhoods to converge in parallel.
This local behaviour is captured in \figurename~\ref{fig:range_impact} which reports the average distance between a network change and a node whose bidder changes its claim in response.
Distances are reported in hops.
A node that changes its demand or gains/loses a neighbour has distance zero.
Neighbours of this node have distance one, and so on.
Range of impact is reported for the six types of change evaluated in Sections~\ref{sec:results_conv_dmnd_chg} and \ref{sec:results_conv_top_chg}.
Each type of change is simulated in 1000 network scenarios, 250 of each traffic load.
The range of impact is less than 1.75 hops on average.

\subsection{Performance with Node Mobility}
\label{sec:results_mobility}

Section~\ref{sec:results_conv_top_chg} addresses the robustness of ATLAS to single topology changes.
We now evaluate its performance in networks with continuous mobility which may not have the opportunity to converge on the TLA allocation.

\begin{figure*}%
\centering
	\subfloat[Relative persistence error.]{
		\includegraphics[width=\subFigureWidth]%
		{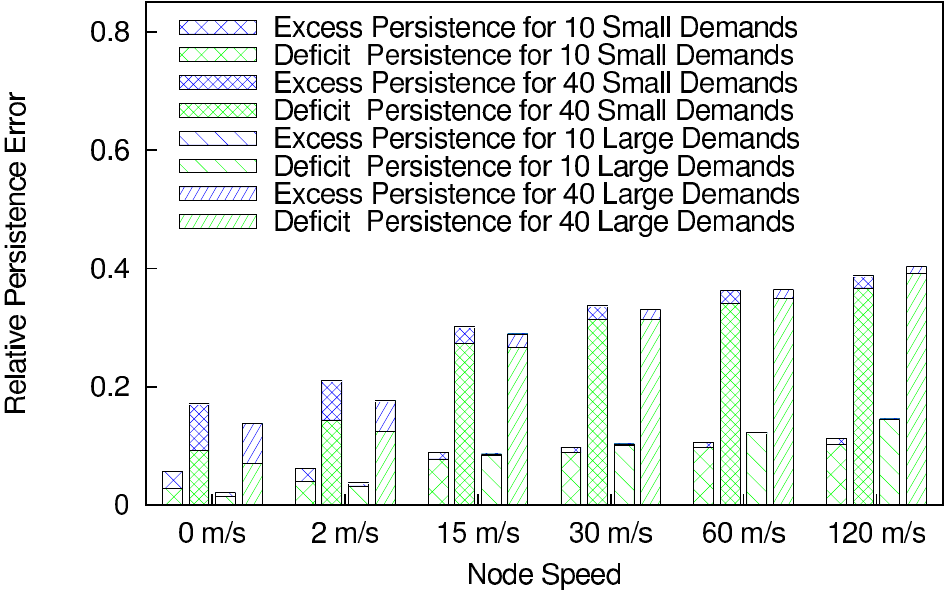}%
		\label{fig:error_top_rwp_mobility}
	}
	\hspace{0.65in}
	\subfloat[Total MAC throughput. ]{
		\includegraphics[width=\subFigureWidth]%
		{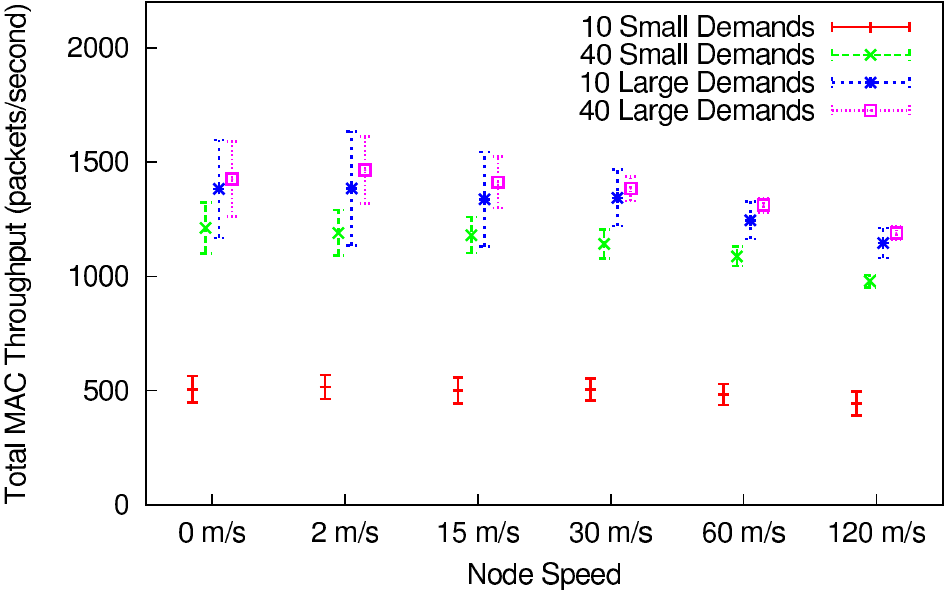}%
		\label{fig:thput_top_rwp_mobility}
	}
	\caption{Relative persistence error and total MAC throughput for varying levels of node mobility.}
	\label{fig:mobility_error_thput}
\end{figure*}

\figurename~\ref{fig:error_top_rwp_mobility} reports persistence error for node speeds ranging from 0~m/s to 120~m/s with 200 scenarios simulated for each node speed, 50 of each traffic load.
Node movements are generated using the steady-state mobility model generator of \cite{mobgen} with a pause time of zero. 
Simulations are run for 20s.
As node speeds increase, so do deficit persistence errors.
The larger deficit errors are an artifact of lost neighbour detection which is delayed by $\nbrTimeOut=0.5$s.
As a result, nodes tend to think their neighbourhoods are more crowded than they are, a tendency that  gets worse as node speeds increase.
In terms of REACT, auctioneers and bidders unnecessarily constrain their offers and claims to accommodate lost neighbours.
The deficit persistences translate to degraded throughput.
\figurename~\ref{fig:thput_top_rwp_mobility} reports MAC throughput for the simulations of \figurename~\ref{fig:error_top_rwp_mobility}.
Even with node speeds of 120~m/s where a node travels its transmission range in 2.1s, throughput degrades modestly, decreasing by less than 20\% compared to static networks.

\begin{figure*}%
\centering
	\subfloat[Relative persistence error.]{
		\includegraphics[width=\subFigureWidth]%
		{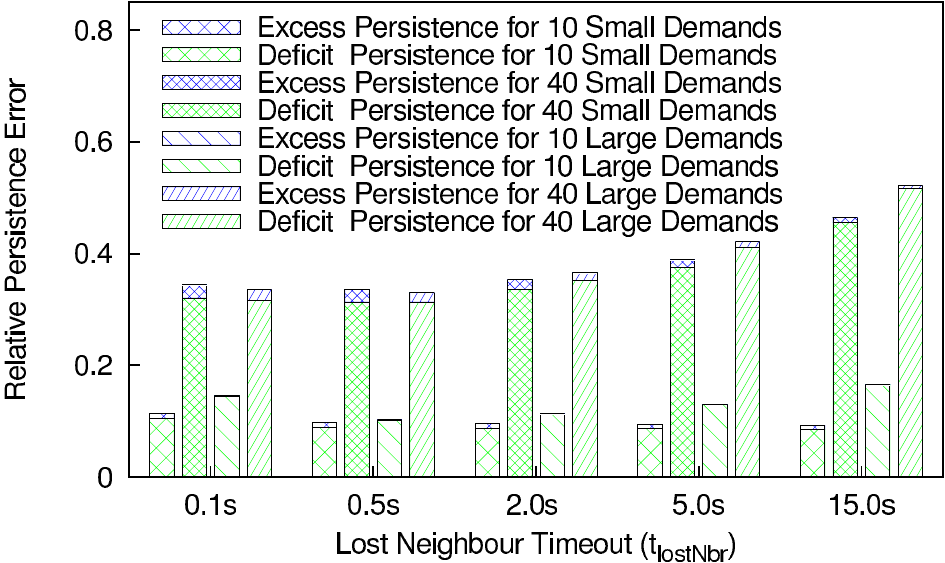}%
		\label{fig:error_top_rwp_timeout}
	}
	\hspace{0.65in}
	\subfloat[Total MAC throughput. ]{
		\includegraphics[width=\subFigureWidth]%
		{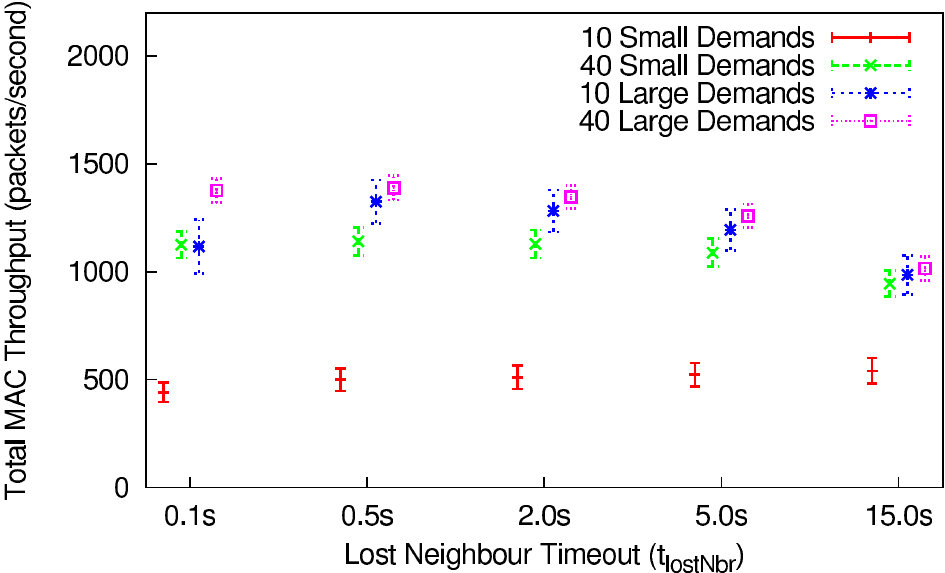}%
		\label{fig:thput_top_rwp_timeout}
	}
	\caption{Relative persistence error and total MAC throughput for varying neighbour timeouts.}
	\label{fig:timeout_error_thput}
\end{figure*}

\figurename~\ref{fig:timeout_error_thput} shows that a large \nbrTimeOut{} exacerbates deficit persistence error and further degrades throughput.
Data is collected from 200 scenarios, 50 of each traffic load.
Each scenario is simulated five times with neighbour timeouts ranging from 0.1s to 15.0s.
Node speeds are fixed at 30~m/s.
Degraded performance is observed for large timeouts, $\nbrTimeOut{} \geq 0.5$s, but also for small timeouts, $\nbrTimeOut{} = 0.1$s.
In networks loaded with 10 large demands, $\nbrTimeOut{} = 0.1$s causes nodes to falsely identify lost neighbours that must be rediscovered.
The remaining simulations are run with $\nbrTimeOut{} = 0.5$s.

\begin{figure}[t]
\centering
\includegraphics[width=\subFigureWidth]
{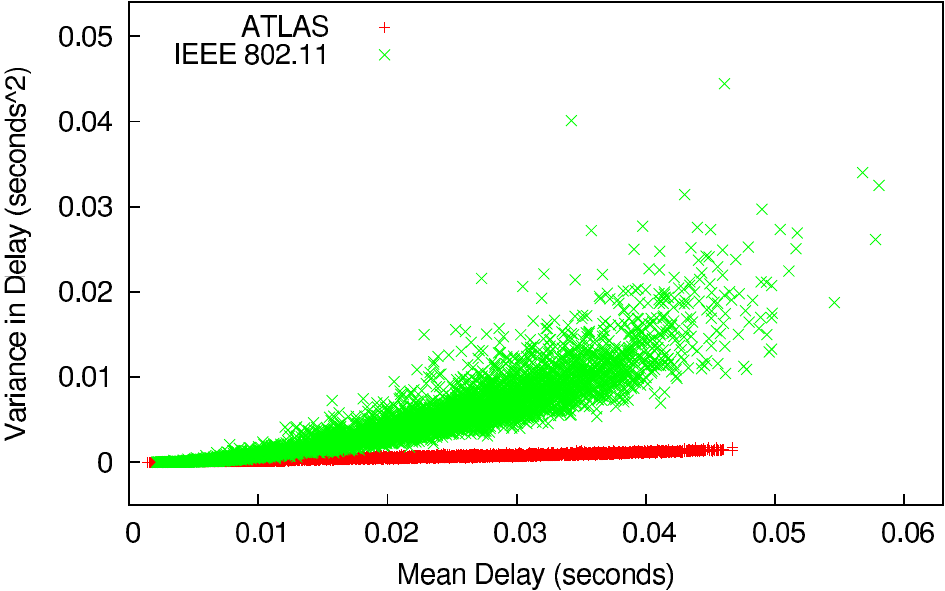}
\caption{
Delays for ATLAS and IEEE 802.11 with node speeds of 30~m/s. 
}
\label{fig:delay_scatter}
\end{figure}

\figurename~\ref{fig:delay_scatter} reports packet delay for ATLAS and IEEE 802.11 for the 200 network scenarios of \figurename~\ref{fig:mobility_error_thput} with node speeds equal to 30~m/s.
IEEE 802.11 is configured with a maximum packet retry count of seven for RTS, CTS, and ACKs and four for data packets \cite{IEEE_802.11}, a mini-slot length of 20$\mu$s, and minimum and maximum contention window sizes of 32 and 1024 slots, respectively.
Each point in the scatter plot reports the average packet delay ($x$-coordinate) and variation in packet delay ($y$-coordinate) for a single node. 
The largest reported average delay is 0.047s for ATLAS and 0.058s for IEEE 802.11. 
The largest reported variation in delay for ATLAS is 0.0016s$^2$, just 3.6\% of the 0.0444s$^2$ reported for IEEE 802.11. 
This impressive reduction in delay variance is crucial to the support of TCP, which we evaluate next.

\subsection{Multi-hop TCP Flows}
\label{sec:results_tcp}

To this point, we have used MAC layer traffic to simulate a diverse set of network scenarios.
We now evaluate the performance of ATLAS using multi-hop TCP flows.
To accommodate the dynamic nature of these flows, each node estimates its own demand by monitoring queue behaviour.
Demand is estimated as the sum of two parts: $w_{\text{enqueue}}$ and $w_{\text{level}}$.
$w_{\text{enqueue}}$ is the percentage of channel required to keep up with the current enqueue packet rate, $w_{\text{enqueue}} = (\text{packet enqueue rate})\times(\text{slot length})$.
$w_{\text{level}}$ is the percentage of channel required to transmit all packets in the queue within 0.2s (\ie 25 slots), $w_{\text{level}} = [(\text{\# packets in queue}) / 0.02\text{s}]\times(\text{slot length})$.

\newcommand{\labelHeight}{1.2in}
\newcommand{\leftLabelWidth}{0.05\textwidth}
\newcommand{\rightLabelWidth}{0.03\textwidth}
\newcommand{\leftfigureWidth}{0.333\textwidth}
\newcommand{\keyWidth}{0.6\textwidth}
\newcommand{\vertSpace}{0.25cm}

\newcommand{\mainTableWidth}{0.98\textwidth}
\newcommand{\leftColWidth}{0.02\textwidth}
\newcommand{\smfigureWidth}{0.3\textwidth}

\begin{figure*}[ht!]
\centering

\begin{tabular}{ >{\centering\arraybackslash} m{\leftColWidth} >{\centering\arraybackslash} m{\mainTableWidth} }

	\begin{turn}{90}
		$y$-axis shows \% of TCP flows achieving minimum required throughput.
	\end{turn}

&

	\begin{minipage}{\mainTableWidth}
	
		\begin{minipage}{\leftLabelWidth}
			\centering
		\end{minipage}
		\begin{minipage}{\smfigureWidth}
			\centering
			\begin{minipage}{\textwidth}\end{minipage}
		\end{minipage}
		\begin{minipage}{\smfigureWidth}
			\centering
			\begin{minipage}{\textwidth}\end{minipage}
		\end{minipage}
		\begin{minipage}{\smfigureWidth}
			\flushright
			\includegraphics[width=\keyWidth]
			{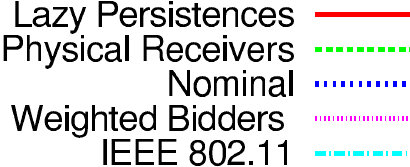}
		\end{minipage}
		\begin{minipage}{\rightLabelWidth}
		\end{minipage}
		
		\vspace{\vertSpace}
		
		\begin{minipage}{\leftLabelWidth}
		\end{minipage}
		\begin{minipage}{\smfigureWidth}
			\centering
			Networks with 2 Flows
		\end{minipage}
		\begin{minipage}{\smfigureWidth}
			\centering
			Networks with 8 Flows
		\end{minipage}
		\begin{minipage}{\smfigureWidth}
			\centering
			Networks with 25 Flows
		\end{minipage}
		\begin{minipage}{\rightLabelWidth}
		\end{minipage}
		
		\vspace{\vertSpace}

		\begin{minipage}{\smfigureWidth}
			\centering
			\includegraphics[width=\textwidth]
			{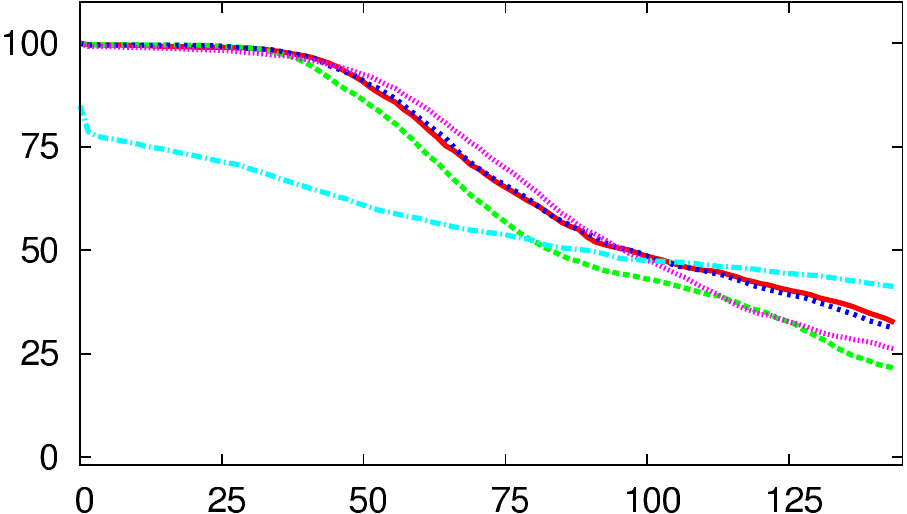}
		\end{minipage}
		\begin{minipage}{\smfigureWidth}
			\centering
			\includegraphics[width=\textwidth]
			{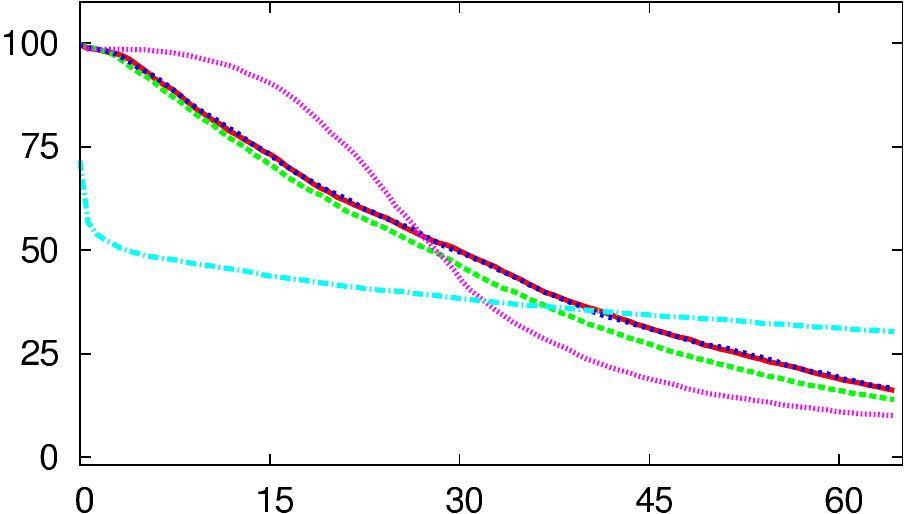}
		\end{minipage}
		\begin{minipage}{\smfigureWidth}
			\centering
			\includegraphics[width=\textwidth]
			{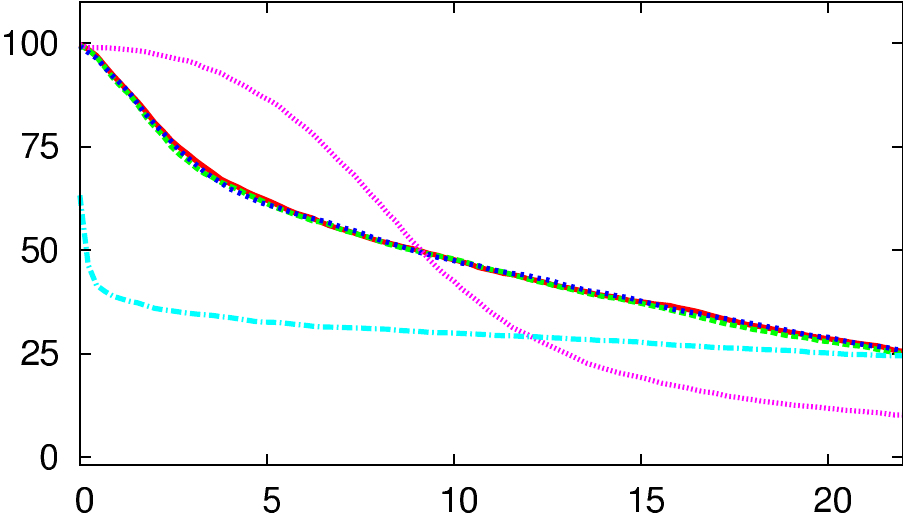}
		\end{minipage}
		\begin{minipage}{\rightLabelWidth}
			\centering
			\begin{turn}{-90}
			\begin{minipage}{\labelHeight}	
			\centering
			All flows
			\end{minipage}
			\end{turn}
		\end{minipage}

		\begin{minipage}{\smfigureWidth}
			\centering
			\includegraphics[width=\textwidth]
			{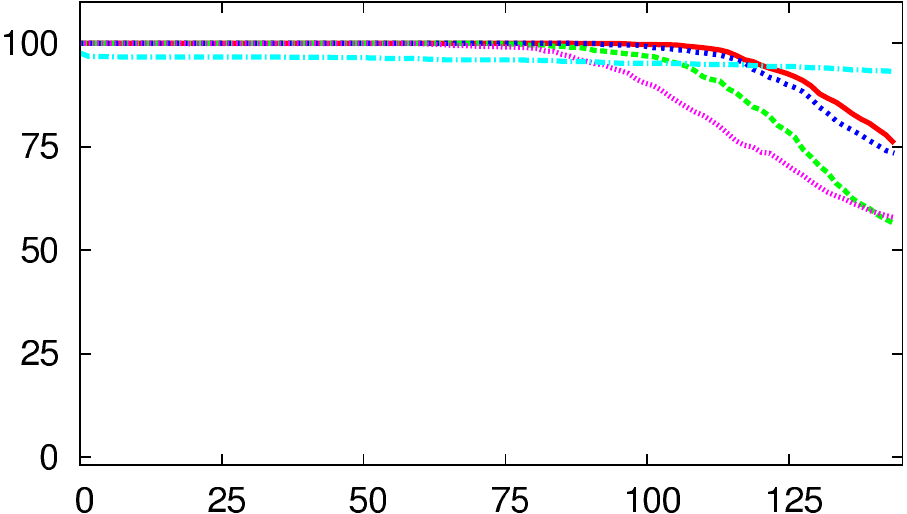}
		\end{minipage}
		\begin{minipage}{\smfigureWidth}
			\centering
			\includegraphics[width=\textwidth]
			{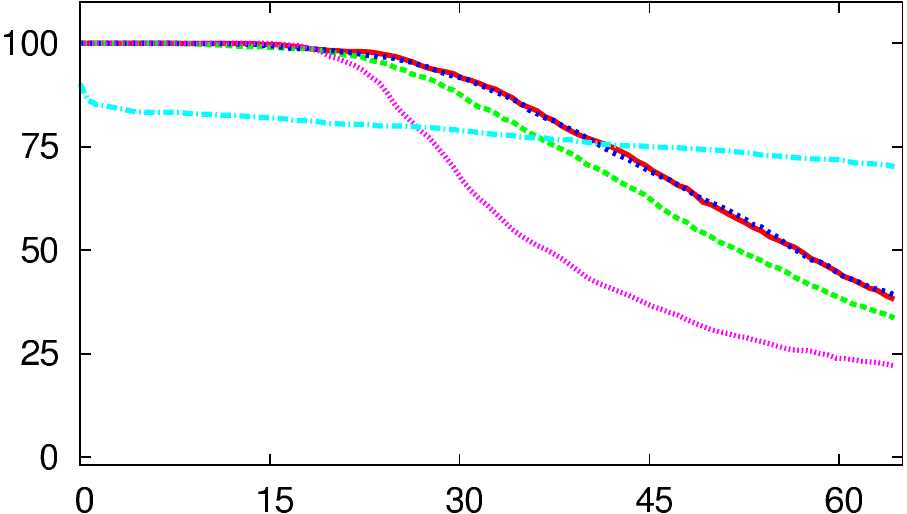}
		\end{minipage}
		\begin{minipage}{\smfigureWidth}
			\centering
			\includegraphics[width=\textwidth]
			{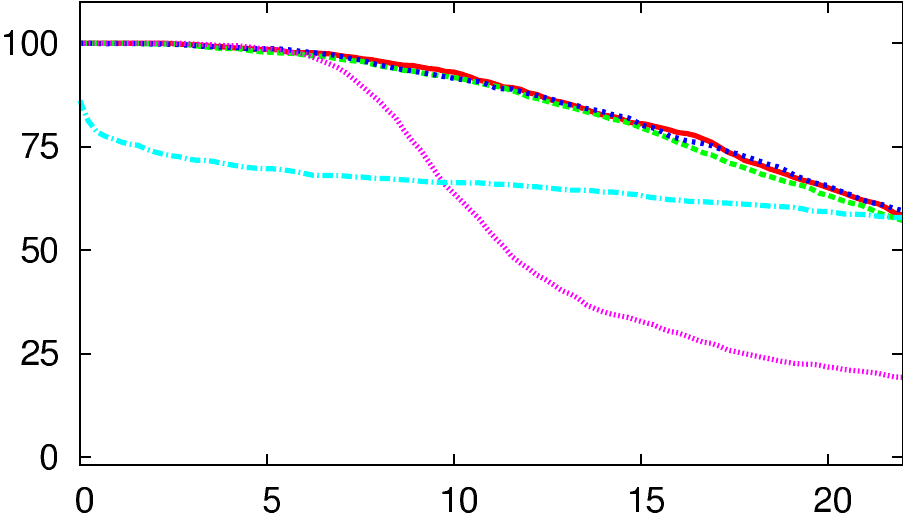}
		\end{minipage}
		\begin{minipage}{\rightLabelWidth}
			\centering
			\begin{turn}{-90}
			\begin{minipage}{\labelHeight}	
			\centering
			1-hop flows
			\end{minipage}
			\end{turn}
		\end{minipage}

		\begin{minipage}{\smfigureWidth}
			\centering
			\includegraphics[width=\textwidth]
			{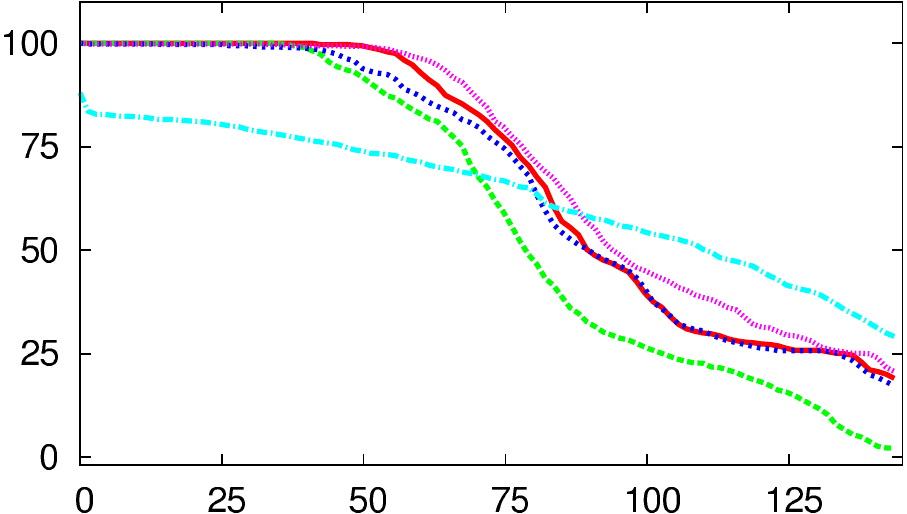}
		\end{minipage}
		\begin{minipage}{\smfigureWidth}
			\centering
			\includegraphics[width=\textwidth]
			{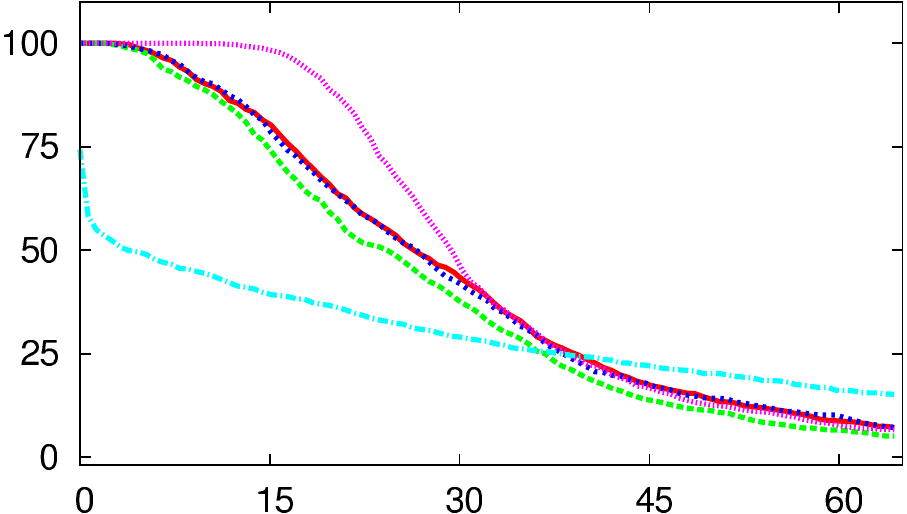}
		\end{minipage}
		\begin{minipage}{\smfigureWidth}
			\centering
			\includegraphics[width=\textwidth]
			{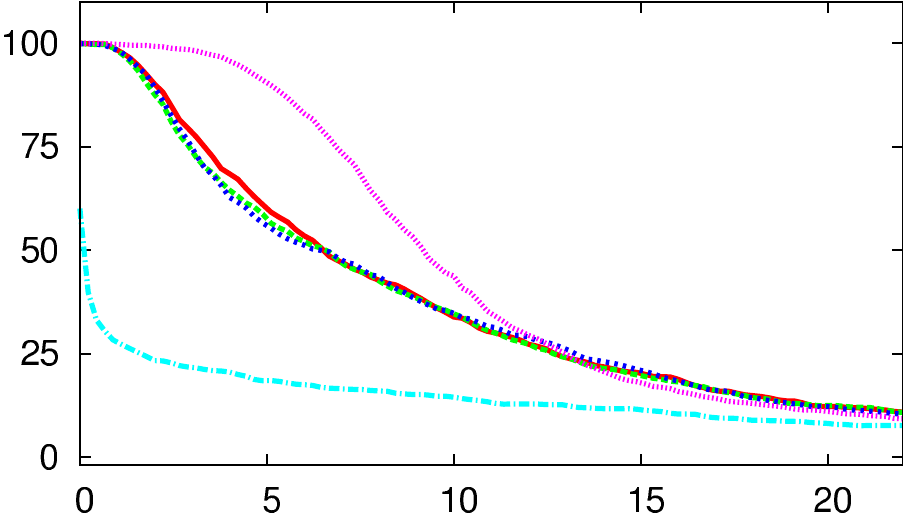}
		\end{minipage}
		\begin{minipage}{\rightLabelWidth}
			\centering
			\begin{turn}{-90}
			\begin{minipage}{\labelHeight}	
			\centering
			2-hop flows
			\end{minipage}
			\end{turn}
		\end{minipage}

		\begin{minipage}{\smfigureWidth}
			\centering
			\includegraphics[width=\textwidth]
			{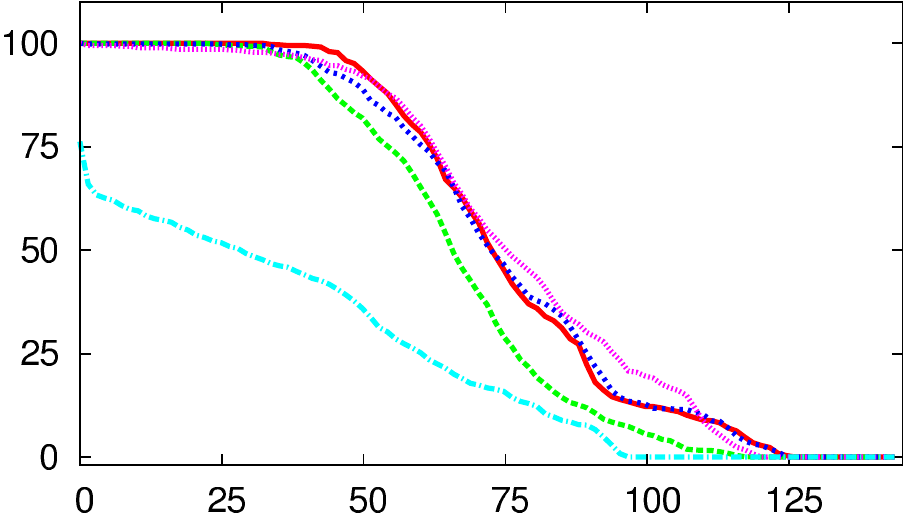}
		\end{minipage}
		\begin{minipage}{\smfigureWidth}
			\centering
			\includegraphics[width=\textwidth]
			{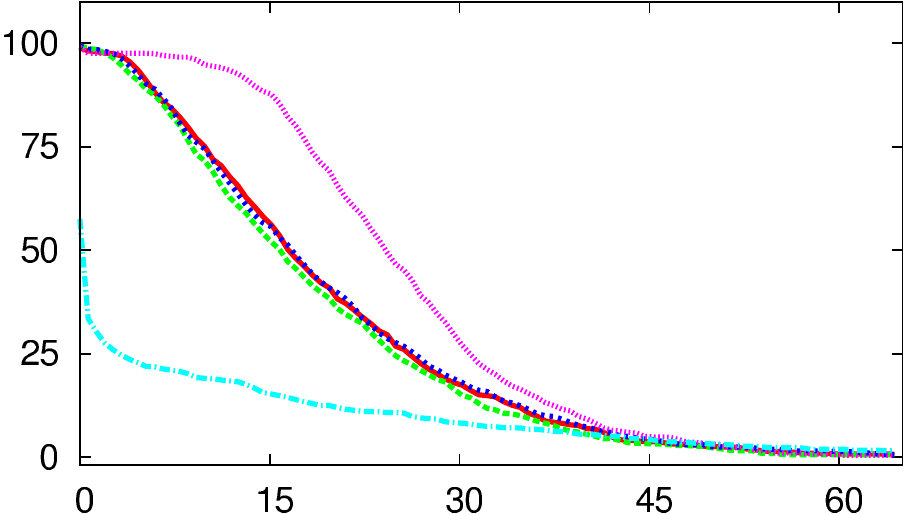}
		\end{minipage}
		\begin{minipage}{\smfigureWidth}
			\centering
			\includegraphics[width=\textwidth]
			{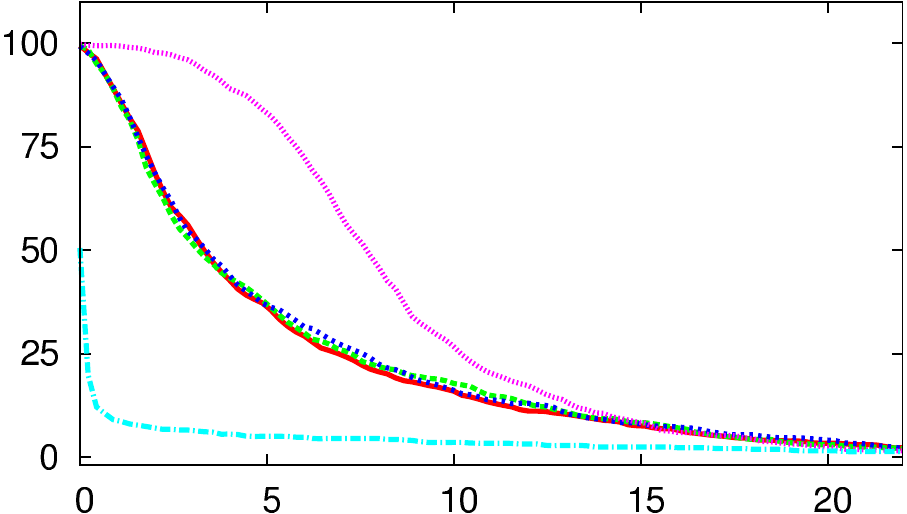}
		\end{minipage}
		\begin{minipage}{\rightLabelWidth}
			\centering
			\begin{turn}{-90}
			\begin{minipage}{\labelHeight}	
			\centering
			3-hop flows
			\end{minipage}
			\end{turn}
		\end{minipage}

		\begin{minipage}{\smfigureWidth}
			\centering
			\includegraphics[width=\textwidth]
			{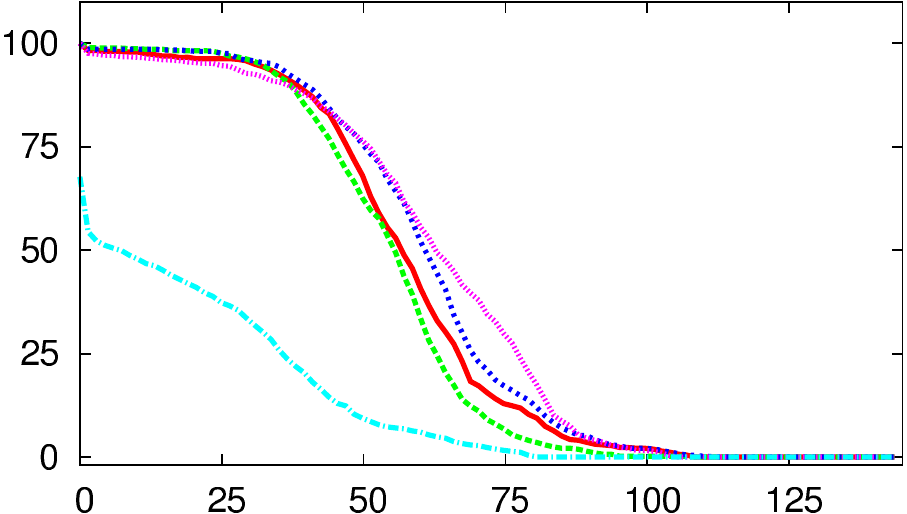}
		\end{minipage}
		\begin{minipage}{\smfigureWidth}
			\centering
			\includegraphics[width=\textwidth]
			{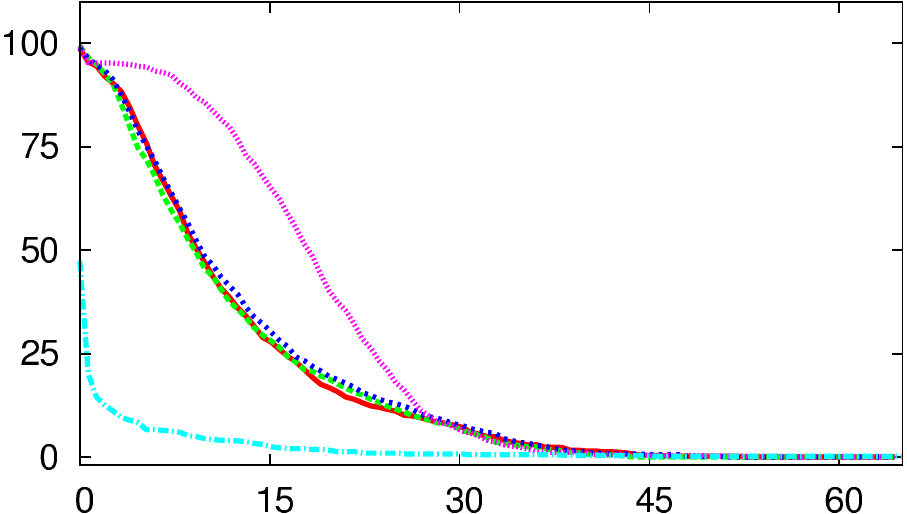}
		\end{minipage}
		\begin{minipage}{\smfigureWidth}
			\centering
			\includegraphics[width=\textwidth]
			{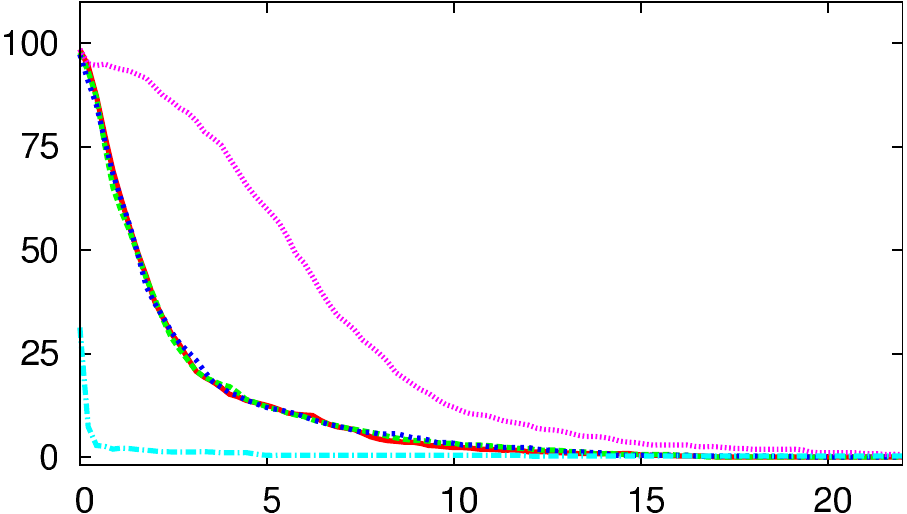}
		\end{minipage}
		\begin{minipage}{\rightLabelWidth}
			\centering
			\begin{turn}{-90}
			\begin{minipage}{\labelHeight}	
			\centering
			4- and 5-hop flows
			\end{minipage}
			\end{turn}
		\end{minipage}
		
		\vspace{\vertSpace}

	\end{minipage}

	\begin{centering}
	$x$-axis shows minimum required throughput for TCP flows in packets/second.
	\end{centering}

\end{tabular}

\caption{
Percent of TCP flows ($y$-axis) achieving a minimum throughput ($x$-axis).
Plots in the left, center, and right columns report on flows from simulations of 2, 8, and 25 flows, respectively.
The plots in the top row report on all flows, regardless of hop count.
Plots in the second, third, and fourth rows report on 1-hop, 2-hop, and 3-hop flows, respectively.
Plots in the fifth row report on 4- and 5-hop flows.
}
\label{fig:tcp_thput}
\end{figure*}

To avoid cross-layer interactions between the MAC and routing protocols, Dijkstra's shortest path algorithm \cite{ATanenbaum_1.03} using accurate knowledge of the global topology computes the next hop address for all packet transmissions.
FTP agents emulate transfer of infinite size files to create flows with throughput limited only by the performance of the network.
Transfers start at time zero and run for 20s.
Nodes are statically placed at random locations in a $300\times1500\text{m}^2$ simulation area.
The source and destination nodes for each file transfer are selected at random.
Each FTP transfer is transported over TCP Reno configured for selective acknowledgements, the extensions of RFC 1323 \cite{RFC_1323}, and 900 byte TCP segments.
The return ACKs are not combined with each other or with other data packets. 
Consequently, the transmission of a single 40-byte TCP ACK consumes an entire transmission slot in ATLAS.
The maximum congestion window size is 32 packets.
Network scenarios are simulated for three traffic loads: networks with 2, 8, and 25 TCP flows.
The number of replicates per traffic load are chosen so that 3000 TCP flows are simulated for each.
Fifteen hundred scenarios are simulated with two TCP flows, 375 with eight TCP flows, and 120 with 25 TCP flows.

We simulate TCP traffic on five MAC protocols: the four configurations of ATLAS and IEEE 802.11.
The configurations of ATLAS use $\persistDefault=0.05$, $\nbrTimeOut=0.5$s, and $p_{\text{min}}=0.01$.
IEEE 802.11 parameters match those described in Section~\ref{sec:results_mobility}.
Each node dynamically sets its bidder weight to one or the number of outgoing TCP flows it services, whichever is larger.

The 15 sub-plots in \figurename~\ref{fig:tcp_thput} show the percentage of flows ($y$-axis) achieving a minimum throughput ($x$-axis).
The distinguishing characteristics of the three unweighted ATLAS configurations are seen in the throughput curves for networks with two flows.
These networks are loaded lightly enough for the auctions at non-receiver nodes to make a difference in the allocation, improving throughput for 2- and 4-hop flows.
These networks also demonstrate how the longer initial packet delays of the Lazy Persistences configuration increase round trip time for 4- and 5-hop flows, preventing TCP from achieving its best throughput.

The Weighted Bidders configuration performs well for multi-hop flows in networks with eight and 25 flows by allocating more to multi-hop flows at the expense of single-hop flows.
Because one-hop flows tend to achieve higher throughput, the configuration maintains a tighter variation in flow throughputs as indicated by the steeper slope of the Weighted Bidders curve in the top right plot of \figurename~\ref{fig:tcp_thput}.

Regardless of configuration, ATLAS surpasses IEEE 802.11 in support of concurrent multi-hop flows.
The interaction between the IEEE 802.11 back-off algorithm and TCP's congestion control is well known \cite{MGerla_01.99}.
In testbed experiments, a single TCP flow with no competition has difficulty reaching a destination four hops away \cite{DKoutsonikolas_01.07}.
Our simulations corroborate these findings, as approximately 50\% of the 4- and 5-hop flows report a throughput of zero.
For networks with 25 demands, nearly 75\% of 2-hop flows are non-functional; 3-, 4-, and 5-hop flows are almost completely shut out.
The throughput of ATLAS is achieved in spite of channel wasted transmitting 40 byte TCP ACKs in their own slots.

\subsection{Comparison with other Scheduled MAC Protocols}
\label{sec:mac_comparison}

Here, we compare the adaptation of ATLAS with several other scheduled protocols including DRAND, Z-MAC, FPRP, and SEEDEX.
Although the first three compute conflict-free schedules, an NP-hard problem \cite{EvenGMT84}, a comparison highlights the agility of ATLAS.

\subsubsection{Adaptation to Topology Changes}
For the simulations of Section~\ref{sec:results_mobility}, the number of neighbour changes (\ie gained or lost neighbours) per second experienced by a node is correlated to the node speed.
When the nodes move at 30 m/s, each node is expected to gain, or lose, a neighbour 2.21 times per second; within 6.3s, the number of neighbour changes is expected to exceed the neighbourhood size.

Based on the run times reported in \cite[Fig.~10]{RheeWMX09}, we estimate DRAND to compute schedules for the networks in Section~\ref{sec:results_mobility} in approximately 4.9s (adjusting for data rate and a two-hop neighbourhood size of 27).
In this time, the topology changes caused by nodes moving at 30~m/s are expected to invalidate the computed schedule.
Z-MAC has the same limitation and, although it compensates by running CSMA/CA to resolve collisions, it does not benefit from its TDMA schedule when nodes are mobile.
In \cite{RheeWMX09}, the run times reported for FPRP schedule generation are comparable to DRAND.
For SEEDEX, nodes discover their two-hop neighbours using a fan-in/fan-out procedure described in \cite{RRozovsky_1.01}. 
However, a practical integration of the procedure into the MAC protocol is not described or evaluated, preventing a comparison of its agility with other MAC protocols.
In contrast to the slow schedule computation times of DRAND, Z-MAC, and FPRP, ATLAS is shown to handle node speeds of up to 120~m/s with only moderate degradation to MAC throughput.

\subsubsection{Adaptation to Changes in Traffic Load}
The persistences achieved by DRAND and SEEDEX are dependent on topology alone; neither adapts to traffic load.
Although Z-MAC adapts to load, it does so by deviating from its underlying schedule, which does not adapt.
FPRP can adapt to load by scheduling a variable number of slots per node; this is done at the expense of both longer frame lengths and longer run times for schedule computation. 
In contrast, ATLAS adapts to traffic load, responding quickly enough to establish and maintain multi-hop TCP flows.

\subsubsection{Continuous Adaptation}
Common to the scheduled schemes mentioned here is the use of a distinct phase for schedule computation (or neighbour discovery for SEEDEX). 
The schedules must be updated in order for the MAC to adapt.
Any fixed period between schedule updates must be selected a priori; it cannot be adjusted for variations in network mobility.
If schedules are to be updated when needed, a mechanism is required to trigger the schedule update. 
This coordination, by itself, is a challenge in an ad hoc network.
In contrast, ATLAS does not employ a schedule computation (or a neighbour discovery) phase and adapts continuously to changes in both topology and traffic load.

\section{Discussion}
\label{sec:discussion}

In this section we discuss open issues and suggest potential applications for REACT and ATLAS.

\subsection{Improved Reliable Transport}

TCP's congestion control algorithm is known to suffer cross-layer interactions with binary exponential back-off (BEB) employed by IEEE 802.11 \cite{MGerla_01.99}.
BEB is short term unfair, allowing a single node to capture the channel at the expense of its neighbours \cite{VBharghavan_1.94,JHastad_01.87} causing high variation in packet delay and making it difficult for TCP to estimate round-trip delay.
Many modifications have been proposed to improve TCP performance over wireless networks \cite{KLeung_01.06}; common approaches are detection of packet loss (differentiating it from congestion) and improved estimation of round trip time.
An alternative is to minimize packet loss and control variation in packet delay at the MAC layer.
ATLAS demonstrates a remarkable control of variation in delay (\figurename~\ref{fig:delay_scatter}) enabling TCP to reliably support 3-, 4-, and 5-hop flows over heavily loaded networks (\figurename~\ref{fig:tcp_thput}).
However, TCP throughput still degrades considerably as the number of hops grows.
Potential areas for future work include the integration of ATLAS into a cross-layer solution for reliable transport over wireless networks and the use of REACT to inform TCP's congestion window size.

\subsection{Selection of Configurable Parameters}

ATLAS has three configurable parameters: \persistDefault, \nbrTimeOut, and \persistMin.
Based on our simulations, [0.01--0.2] is an acceptable range for \persistDefault{} (\figurename~\ref{fig:boot_conv_defaultp}) and [0.1s--2s] is an acceptable range for \nbrTimeOut{} (Figs.~\ref{fig:error_top_rwp_timeout}, \ref{fig:thput_top_rwp_timeout}).
In \cite{JLutz_01.12}, \persistMin$=$0.1s is found to be acceptable for a protocol that enforces \persistMin{} at all nodes and at all times.
Because ATLAS employs \persistMin{} temporarily, and only when needed, it is less sensitive to the selection of \persistMin.
Although results show ATLAS to be robust to parameter selection, tuning may be required in other scenarios or in a hardware implementation.

\subsection{Dynamic Selection of Auction Capacity}

ATLAS targets 100\% channel allocation by setting auction capacities in REACT to one.
Although simulation results show this to be an adequate choice, it is not clear whether performance can be improved by under- or over-allocating the channel.
Indeed, optimal auction capacities (however optimal is defined) are dependent on network topology and quality of the communication channel.
We leave a thorough analysis of auction capacity selection to future work, pointing out here that REACT adapts continuously, allowing auction capacities to be adjusted dynamically, if necessary.


\subsection{Potential Applications for REACT}

The weighted TLA allocation opens doors for several potential uses.
In the simulations of Section~\ref{sec:results_tcp}, a bidder's weight is set according to the number of flows it services.
It may be desirable to set weights according to queue levels, demand magnitudes, neighbourhood sizes, node betweenness \cite{LFreeman_01.77}, distance from a point of interest (\ie an access point or a common sink), position in a multicast/broadcast tree, or path hop count.
The key observation is that ATLAS maintains flexibility by allowing nodes to define bidder weights arbitrarily to suit the needs of the network.

While computation of persistences is the primary motivation for this work, REACT is not limited to this purpose.
Consider the Physical Receivers configuration with node demands set to one.
The resulting allocation is independent of actions taken by the upper network layers and, therefore, can  inform decisions made by those layers.
It can serve as a measure of potential network congestion---small allocations are assigned in dense neighbourhoods containing many potentially active neighbours.
The routing protocol can use the allocation to discover alternate routes around congestion.

An intriguing application is the implementation of differentiated service at the MAC layer.
IEEE 802.11e \cite{IEEE_802.11e} enhances the distributed coordination function by implementing four access categories; 
an instance of the back-off algorithm is run per access category, each with its own queue.
The probability of transmission of each access category is manipulated independently through selection of contention window size and inter-frame space.
This permits higher priority traffic to capture the channel from lower priority traffic.

Similar results can be achieved by four instances of REACT, each computing the allocation for a single access category.
Prioritization is achieved through dynamic coordination of the four auction capacities at each node.
A potential strategy sets the capacity for each access category equal to one minus the allocation to higher priority access categories.
As a result, higher priority auctions are permitted to starve lower priority auctions of capacity, effectively distributing channel access to high priority traffic.
Alternatively, auction capacities can be selected to ensure a minimum or maximum percentage of the channel is offered to an access category.

A network can run multiple instances of REACT.
For example, an instance of the Physical Receivers configuration with all demands set to one can be run concurrently with four instances configured to support differentiated service.
Alternatively, multiple instances of REACT can be used to allocate more than one set of resources concurrently.

\subsection{Assumptions Made by ATLAS}
Two key assumptions are made by ATLAS in its computation of the TLA allocation using REACT:
(1) The offers and claims received by a node are accurate.
(2) The offers and claims of a node are eventually received by all neighbouring nodes.
The first assumption is reasonable, provided received packets are checked for errors by the link layer.
The second assumption is almost certainly invalid; asymmetric communication, interference beyond the range of transmission, and signal fading are common in wireless communication and can prevent the delivery of offers and claims.
Under realistic conditions, REACT may not converge on the TLA allocation, risking over-allocation of the channel.
In practice, auctions can adjust their capacities to mitigate the over-allocation.
Every node knows the persistences of its neighbours (from bidder claims) and can compute the expectation for collisions on the channel.
Significant deviations above this expectation can trigger the auction to lower its capacity.
An evaluation in a testbed of real radios is necessary to understand the sensitivity to anomalies on the wireless channel and the effectiveness of adjusting auction capacities to accommodate channel conditions.

The evaluation of ATLAS in Section~\ref{sec:sim-results} assumes both slot and frame synchronization;
ATLAS does not require either.
The computation of the TLA allocation by REACT does not rely on a frame structure and the expected performance of the random schedules is not affected by loss of frame synchronization. 
Even without slot synchronization, REACT can compute the TLA allocation; however, loss of slot synchronization may reduce channel capacity by 50\% (see Aloha vs. slotted Aloha in \cite{ATanenbaum_1.03}). 
ATLAS can accommodate the lower channel capacity by reducing auction capacity.
This technique may allow ATLAS to be run on commodity IEEE 802.11 hardware \cite{TinnirelloBGGGG2012} that lacks native support for slot synchronization. 
This is a subject of our current research.

\subsection{Enhancing Existing MAC Protocols}

We have used REACT to compute persistences to be employed within ATLAS, a slotted MAC protocol. 
Alternatively, REACT can be run on top of the IEEE 802.11 MAC by embedding claims and offers in the headers of existing control and data messages.
The TLA allocation can be used to inform the selection of contention window sizes, eliminating the need for (and negative side effects of) binary exponential back off.
We are currently working to integrate REACT into IEEE 802.11.
Another alternative (and more ambitious) approach is to implement TLA persistences in a topology-dependent MAC that computes conflict-free schedules.
Only a few topology-dependent schemes allow a node to reserve more than one slot in a frame (\ie \cite{CZhu_1.01,JGentian_1.12}), and those do not define how many slots a node {\em should} reserve. 
The TLA allocation can establish a permissible number of slots to be reserved by each node, given the current topology and traffic load.

\section{Conclusion}
\label{sec:conclusion}

We have proposed REACT, a distributed auction that converges continuously on the TLA allocation, adapting to changes in both topology and traffic load.
The utility of REACT is demonstrated through integration into ATLAS which we simulate under a wide variety of network scenarios.
The results presented suggest that REACT can effectively inform the selection of transmitter persistences, and that ATLAS can provide robust, reliable, and scalable services.
The application of REACT is not restricted to the computation of transmitter persistences.
It has the potential to inform routing and admission control decisions, to enable differentiation of service at the MAC layer, and even to allocate other node resources.
In this context, the REACT algorithm provides a potential solution to the immediate challenge of medium access control, but also shows promise as a tool for use in network protocol design in general.

\section*{Acknowledgement}

The authors appreciate the useful comments provided by the anonymous reviewers.


\bibliographystyle{abbrv}
\bibliography{list}

\begin{IEEEbiography}
[{\includegraphics[width=1in,height=1.25in,trim=0 175 0 25,clip,keepaspectratio]{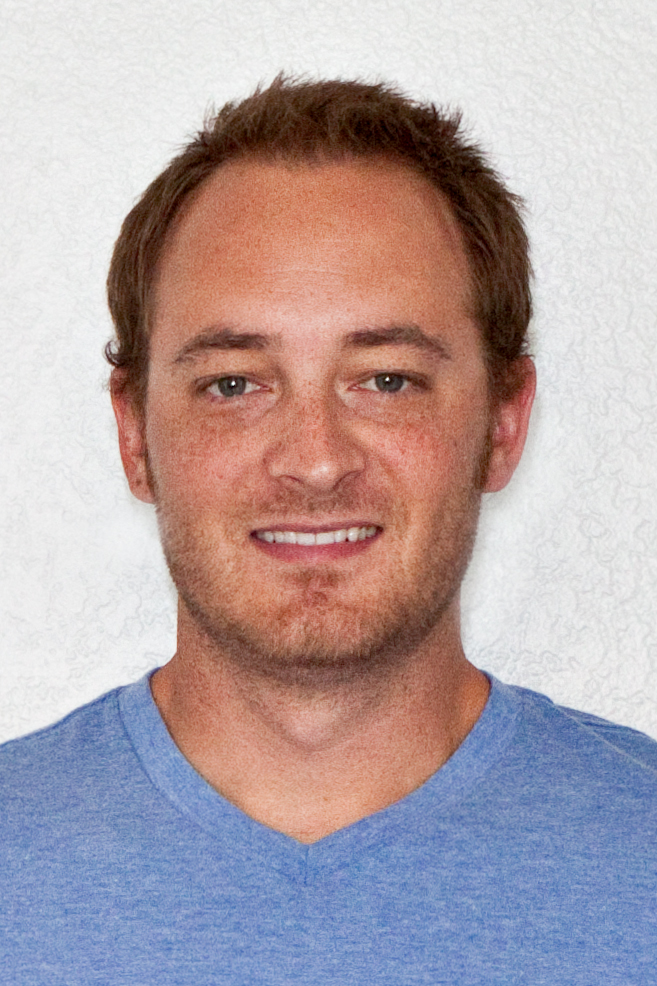}}]
{Jonathan Lutz}
earned his B.S. in Electrical Engineering from Arizona State University, Tempe, Arizona, in 2000 and his M.S. in Computer Engineering from the University of Waterloo, Waterloo, Canada, in 2003.
He is currently working on his Ph.D. in Computer Science at Arizona State University.
His research interests include medium access control in mobile ad hoc networks.
\end{IEEEbiography}

\begin{IEEEbiography}
[{\includegraphics[width=1in,height=1.25in,trim=0 0 0 0,clip,keepaspectratio]{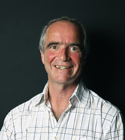}}]
{Charles J.\ Colbourn}
earned his Ph.D. in 1980 from the University of Toronto, and is a Professor of Computer Science and Engineering at Arizona State University.
He is the author of {\em The Combinatorics of Network Reliability} (Oxford), {\em Triple Systems} (Oxford), and 320 refereed journal papers focussing on combinatorial designs and graphs with applications in networking, computing, and communications.
In 2004, he was awarded the Euler Medal for Lifetime Research Achievement by the Institute for Combinatorics and its Applications.
\end{IEEEbiography}

\begin{IEEEbiography}
[{\includegraphics[width=1in,height=1.25in,clip,keepaspectratio]{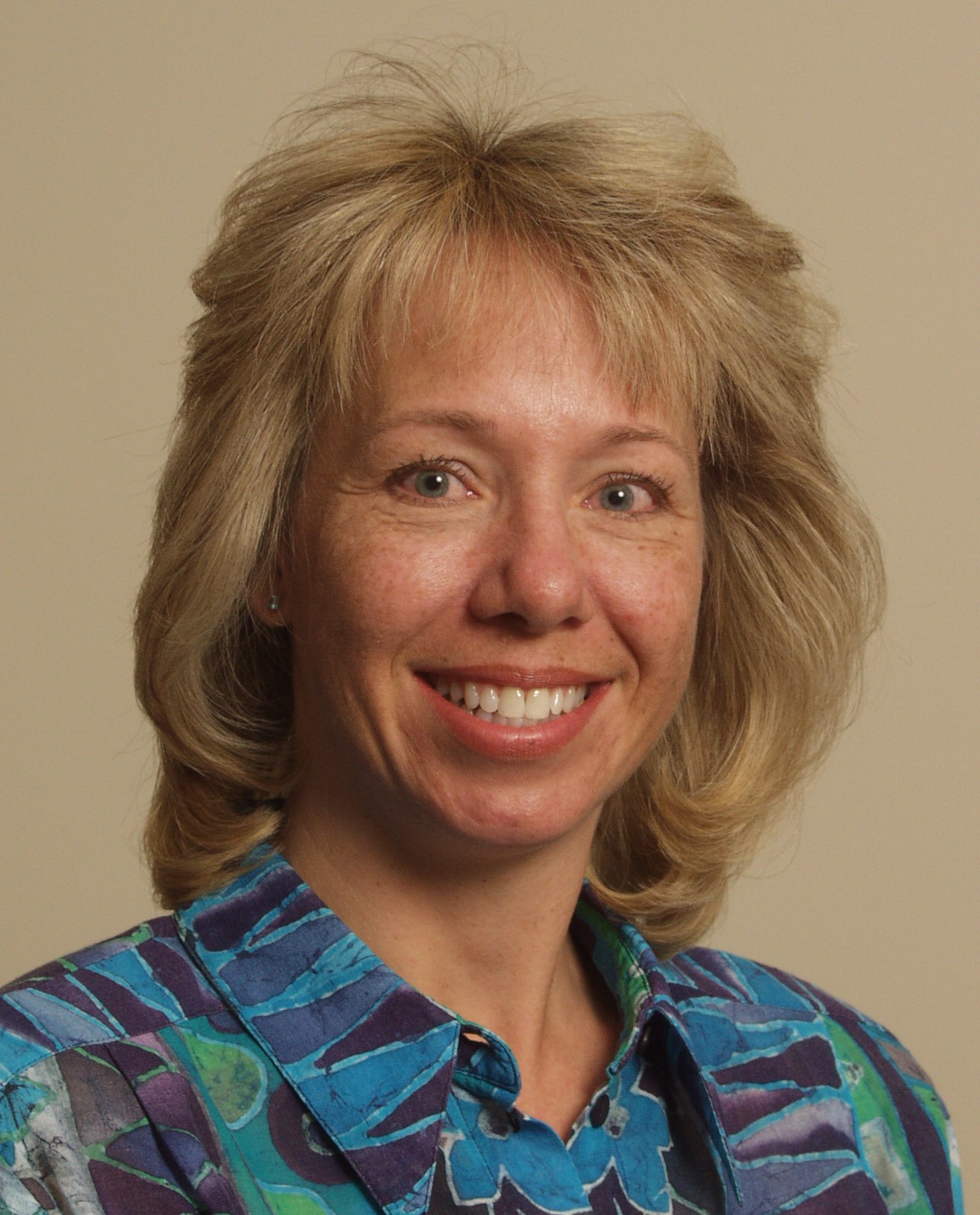}}]
{Violet R.\ Syrotiuk}
earned her Ph.D. in Computer Science from the University of Waterloo (Canada). She is an Associate Professor of Computer Science and Engineering at Arizona State University. Her research has been supported by grants from NSF, ONR, and DSTO, and contracts with LANL, Raytheon, General Dynamics, and ATC. She serves on the editorial boards of Computer Networks and Computer Communications, as well as on the technical program and organizing committees of several major conferences sponsored by ACM and IEEE.
\end{IEEEbiography}

\vfill

\end{document}